\documentclass[11pt]{article}
\usepackage{amssymb,amsmath}
\usepackage{doublespace}
\usepackage{overcite}
\usepackage{epsf}

\begin{document}
\begin{spacing}{1.0}
\title{
ONE-DIMENSIONAL ARRAYS OF OSCILLATORS: ENERGY LOCALIZATION IN
THERMAL EQUILIBRIUM 
}

\author{Ramon Reigada\footnote{Permanent address: 
Departament de Qu\'{\i}mica-F\'{\i}sica, Universitat de Barcelona,
Avda. Diagonal 647, 08028 Barcelona, Spain}\\
Department of Chemistry and Biochemistry 0340\\
University of California San Diego\\
La Jolla, California 92093-0340\\
\and
Aldo H. Romero\footnote{Present address: Max-Planck Institut f\"{u}r
Festk\"{o}rperforschung, Heisenbergstr. 1, 70569 Stuttgart, Germany}\\
Department of Chemistry and Biochemistry 0340\\
and Department of Physics\\
University of California, San Diego\\
La Jolla, California 92093-0340\\
\and
Antonio Sarmiento\footnote{Permanent address: 
Instituto de Astronom\'{\i}a, Apdo. Postal 70-264, Ciudad Universitaria,
M\'{e}xico D. F. 04510, M\'{e}xico}\\
Department of Chemistry and Biochemistry 0340\\
University of California San Diego\\
La Jolla, California 92093-0340\\
\and
Katja Lindenberg\\
Department of Chemistry and Biochemistry 0340\\
and Institute for Nonlinear Science\\
University of California San Diego\\
La Jolla, California 92093-0340}
\date{\today }
\maketitle

\end{spacing}
\begin{spacing}{1.5}
\begin{abstract}
  
All systems in thermal equilibrium exhibit a spatially variable
energy landscape due to thermal fluctuations.  Thus at any instant
there is naturally a thermodynamically driven localization of energy
in parts of the system relative to other parts of the system. 
The specific characteristics of the spatial landscape such as, for example,
the energy variance, depend on the thermodynamic properties of
the system and vary from one system to another. 
The temporal persistence of a given energy landscape, that is, the way
in which energy fluctuations (high or low) decay toward the thermal
mean, depends on the dynamical features of the system.  We discuss the
spatial and temporal characteristics of spontaneous energy localization
in 1D anharmonic chains in thermal equilibrium.

\end{abstract}

\section{Introduction}
\label{intro}
The pioneering work of Fermi, Pasta and Ulam~\cite{FPU} demonstrated
that a periodic lattice of coupled nonlinear oscillators is not
ergodic, and that energy in such a lattice may never be distributed
uniformly. A great deal of work has followed that classic
paper trying to understand
how energy is distributed in discrete nonlinear
systems~\cite{Lepri,Hu,Allen,Bourbonnais,Visco,Willis}. Specifically, the
possibility of spontaneous energy localization in perfect anharmonic
lattices has been a subject of intense
interest~\cite{Dauxois,Szeftel,Tsironis,Bilbault,Brown,Willis,Takeno}.
The existence of solitons and more generally of breathers
and other energy-focusing mechanisms, 
and the stationarity or periodic recurrence or even
slow relaxation of such spatially localized excitations, are viewed
as nonlinear  phenomena with important consequences 
in many physical systems~\cite{Tsironis,Zakharov,Bulsara}. 

The interest in the distribution and motion of energy in perfect
arrays arises in part because localized energy in these systems may
be {\em mobile},
in contrast with systems where energy localization occurs through
disorder.  The interest also
arises because such arrays may themselves serve as models for a heat bath
for other systems connected to them \cite{Sancho}. Albeit in different
contexts, ``perfect" arrays serving as energy storage and transfer
assemblies for chemical or photochemical processes are not uncommon
\cite{Schulten,Scott}.

The study of anharmonic chains and of higher-dimensional discrete
arrays has been less than systematic, certainly an inevitable consequence
of the breadth and mathematical difficulty of the subject.  Some
studies (including the work of
Fermi, Pasta, and Ulam) deal with microcanonical arrays.  Here one
observes the way in which a given constant amount of energy distributes
itself among the elements of the array.  
The notion of ``temperature" usually does not
enter in these discussions, although such an association could be made if
the energy is randomly distributed.  
Other studies of anharmonic chains (far more limited in number)
deal with systems subject to external noise and other external forces.
The questions of interest here involve the ways in which noise can
enhance (as in noise-enhanced signal propagation\cite{Bulsara}) or
even totally modify
(as in noise-induced phase transitions\cite{Chris}) the properties
of the nonlinear
array.  Even more limited has been the study of systems
that are in thermal contact with one or more
external heat baths maintained at a
constant temperature \cite{Tsironis,Brown}.  Here the questions
usually revolve
around the  robustness against thermal fluctuations
of stationary or quasi-stationary solutions of
the microcanonical problem.  In both microcanonical and
canonical systems, some work concentrates
on stationary states or long-time behavior or equilibrium properties of
the array, while other work deals with transport properties or with
the approach to equilibrium.  Furthermore, there is variation 
in the portion of the potential where the nonlinearity resides. 
Thus, in some cases the elements of the array are themselves nonlinear
while in others it is the coupling between elements that is nonlinear (and,
on occasion, both are nonlinear).  

Within this broad setting,
our interest in this paper focuses on one-dimensional arrays of
classical oscillators in {\em thermal equilibrium}~\cite{Brown}. 
An understanding of thermal equilibrium properties and
the effects of nonlinearities on these properties is a prerequisite to the
perhaps more interesting analysis of the nonequilibrium behavior of
anharmonic lattices in the presence of thermal fluctuations and the
approach to equilibrium in such systems.  In particular, here we deal with
the case of ``diagonal anharmonicity," that is, the
nonlinearity in our model is 
inherent within each oscillator in the array (representing, for example,
intramolecular interactions), while the connections
between oscillators (representing, for example, intermolecular
interactions) are ordinary linear springs.  The anharmonicity may be
soft or hard.  We explore the conditions that lead to spontaneous energy
localization in one or a few of the oscillators in the array, and 
the time it takes for a given energy landscape to change to a
different landscape.  One could undertake a parallel study in systems with
anharmonic interactions between oscillators (``off-diagonal" anharmonicity).
We address such systems in subsequent work \cite{Sarm}.

The energy landscape is determined by the local potential of each
oscillator, and by the channels of energy exchange in and out
of each of the
oscillators.  The couplings between oscillators provide
one such exchange channel, and the coupling of the array with the heat bath
provides the other.  We shall see that
different arrays (soft, hard) behave very differently in response to these
channels.  We broadly anticipate our conclusions by revealing
that 1) persistent energy localization occurs in
arrays of weakly coupled soft oscillators even when strongly coupled to
a heat bath (while such localization is absent in the hard chain);
2) persistent localization occurs in strongly coupled hard arrays provided
they are weakly coupled to a heat bath (while such localization is absent
in the soft chain); 3) quasi-dispersionless mobility of localized energy 
requires off-diagonal anharmonicity.

These remarks point to the fact that our analysis of anharmonic
chains in thermal
equilibrium could start from two ``opposite" viewpoints.  On the one hand,
we might start by analyzing uncoupled oscillators in thermal equilibrium and
then proceed to investigate what happens if we couple these oscillators
to one another.  This approach focuses on the entropic localization
mechanism \cite{Brown} and the way in which the coupling between the
oscillators eventually degrades it.  On the other hand,
we might start with a coupled isolated chain, focus on energetic
localization mechanisms in such a chain \cite{Willis},
and then proceed to investigate
the ways in which thermal fluctuations and dissipation affect such
local structures.  Since we are explicitly interested in localization
in the presence of thermal fluctuations, and since entropic effects
have received far less attention than energetic ones, we choose to
follow the former approach.

No matter the sequence of our queries, since here our interest lies
mainly in understanding energy localization in a nonlinear discrete
array in {\em thermal equilibrium} and the way in which
thermal effects depend on system parameters, we pose our questions
as follows: 
\begin{itemize}
\item
How is the energy distributed in an equilibrium nonlinear chain at any
given instant of time, and how does this distribution depend
on the anharmonicity?  In other words, can one talk about {\em spontaneous
energy localization} in thermal equilibrium, and, if so, what are the
mechanisms that lead to it?  
\item
How do local energy fluctuations in such an equilibrium array relax in
a given oscillator? Are there circumstances in the equilibrium system
wherein a given oscillator remains at a high level of excitation for
a long time? 
\item
Can local high-energy fluctuations move in some nondispersive fashion
along the array? In other words, can an array in thermal equilibrium
transmit long-lived high-energy fluctuations (if indeed they exist)
from one region of the array to another without too much energy loss to
dispersion?
\end{itemize}

The answers to these questions have not been found analytically,
and are for that reason most clearly presented in comparative fashion. 
Starting with an ensemble of uncoupled oscillators at thermal equilibrium,
one knows exactly the behavior of a {\em single harmonic oscillator}
and can say a great deal about the behavior of a {\em single anharmonic
oscillator} from general thermodynamic considerations.  Thus, for instance,
the mean energy of a single harmonic oscillator in thermal equilibrium
at temperature $T$ is $E=k_BT$ 
($k_B=$Boltzmann's constant).  This energy is on average divided equally
between kinetic and potential (a partition that enters importantly 
in questions concerning landscape persistence). A simple virial analysis
immediately shows that a soft anharmonic oscillator in thermal
equilibrium has energy {\em greater} than $k_BT$ while a hard
anharmonic oscillator has energy {\em smaller} than
$k_BT$.  Both share the property of the harmonic oscillator that the
average kinetic energy is $k_BT/2$, but their average potential energies
differ.  One also knows exactly the energy fluctuations in a harmonic
oscillator: the energy variance
$\sigma^2$ is equal to $k_B^2T^2$, and the ratio of $\sigma$ to
$E$ is therefore independent of temperature. The energy fluctuations
are easily determined to be greater in a soft oscillator and smaller
in a hard oscillator. From these facts one can arrive at rather definitive
qualitative conclusions regarding the distribution and persistence 
of energy in ensembles of single oscillators and the effects of the
anharmonicities on these features~\cite{Brown}. 

The situation becomes more complicated when such oscillators are
connected to one another.  Not only can the oscillators now
exchange energy with the heat bath, but there are also coupling
channels whereby oscillators can exchange energy with one another.
The interplay of these various energy exchange channels and
the effects of anharmonicity on this interplay are some of the issues to
be addressed in this work.

This paper is organized as follows.
In Section~\ref{model} we introduce our model and notation.  We fix some of
the parameter values and briefly discuss the numerical methods used in our
simulations.   Here we introduce the hard, harmonic, and soft local
potentials to be compared.
In Section~\ref{statloc} we review and illustrate previous  results
for uncoupled oscillators in thermal equilibrium so as to establish the
background for the coupled systems.  The phenomenon of ``entropic
localization," whereby ensembles of single thermalized soft
oscillators localize and retain energy more effectively than harmonic or
hard oscillators, is recalled.
In Section~\ref{coupled} we explore the consequences of coupling our
oscillators. 
In Section~\ref{Transport} we briefly address the mobility of energy
fluctuations in our systems.
Finally, Section~\ref{conclusions} summarizes our findings and
anticipates further studies.

\section{The Model and Numerical Methods}
\label{model}

\begin{figure}[htb]
\begin{center}
\leavevmode
\epsfxsize = 4.6in
\epsffile{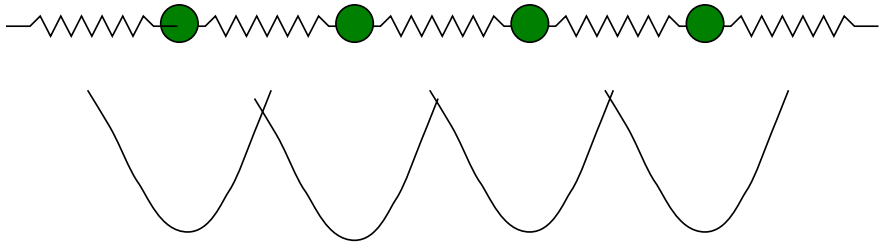}
\vspace{-0.2in}
\end{center}
\caption
{
Illustration of the 1D chain considered in this work.
Each oscillator in the chain experiences an on-site potential
and is harmonically bound to its nearest neighbors.}
\label{fig:cartoon}
\end{figure}

Our system is a one-dimensional chain of N identical unit-mass oscillators
labeled $i=1,2,\cdots,N$ with harmonic nearest-neighbor
interactions and on-site potentials $V(x_i)$ that may be hard,
harmonic or soft. Here $x_i$ is the displacement of oscillator $i$ from its
equilibrium position, with associated momentum $p_i$. We assume periodic
boundary conditions.   The Hamiltonian of the system is
\begin{equation}
H = \sum_{i=1}^N \left( \frac{p_i^2}{2 m} + \frac{1}{2} k(x_i - x_{i+1})^2
+ V(x_i) \right)~,
\label{hamiltonian}
\end{equation}
where $k$ is the intermolecular force constant.
Figure~\ref{fig:cartoon} is a schematic of the  model. 

To represent the thermalization of our chain 
the model is further expanded to
include the Langevin prescription for coupling a system to a heat bath
at temperature $T$ via
fluctuating and dissipative terms. The stochastic equations of motion
for the chain are then given by the Langevin equations 
\begin{equation}
\ddot{x_i}
= -k (2x_i - x_{i+1} - x_{i-1}) - \gamma \dot{x_i}
- \frac{d V(x_i)}{dx_i} + \eta_i(t)
\label{lang}
\end{equation}
where a dot represents a derivative with respect to time.
The $\eta_i(t)$ are mutually uncorrelated
zero-centered Gaussian
$\delta$-correlated fluctuations that satisfy the fluctuation-dissipation
relation:
\begin{equation}
\langle \eta_i(t) \rangle = 0, \;\;\;\;\;\;\;\;\;\;\;\;
\langle \eta_i(t) \eta_j(t') \rangle = 2 \gamma k_B T
\delta_{ij}\delta(t-t')~.
\end{equation}

\begin{figure}[htb]
\begin{center}
\leavevmode
\epsfxsize = 3.in
\epsffile{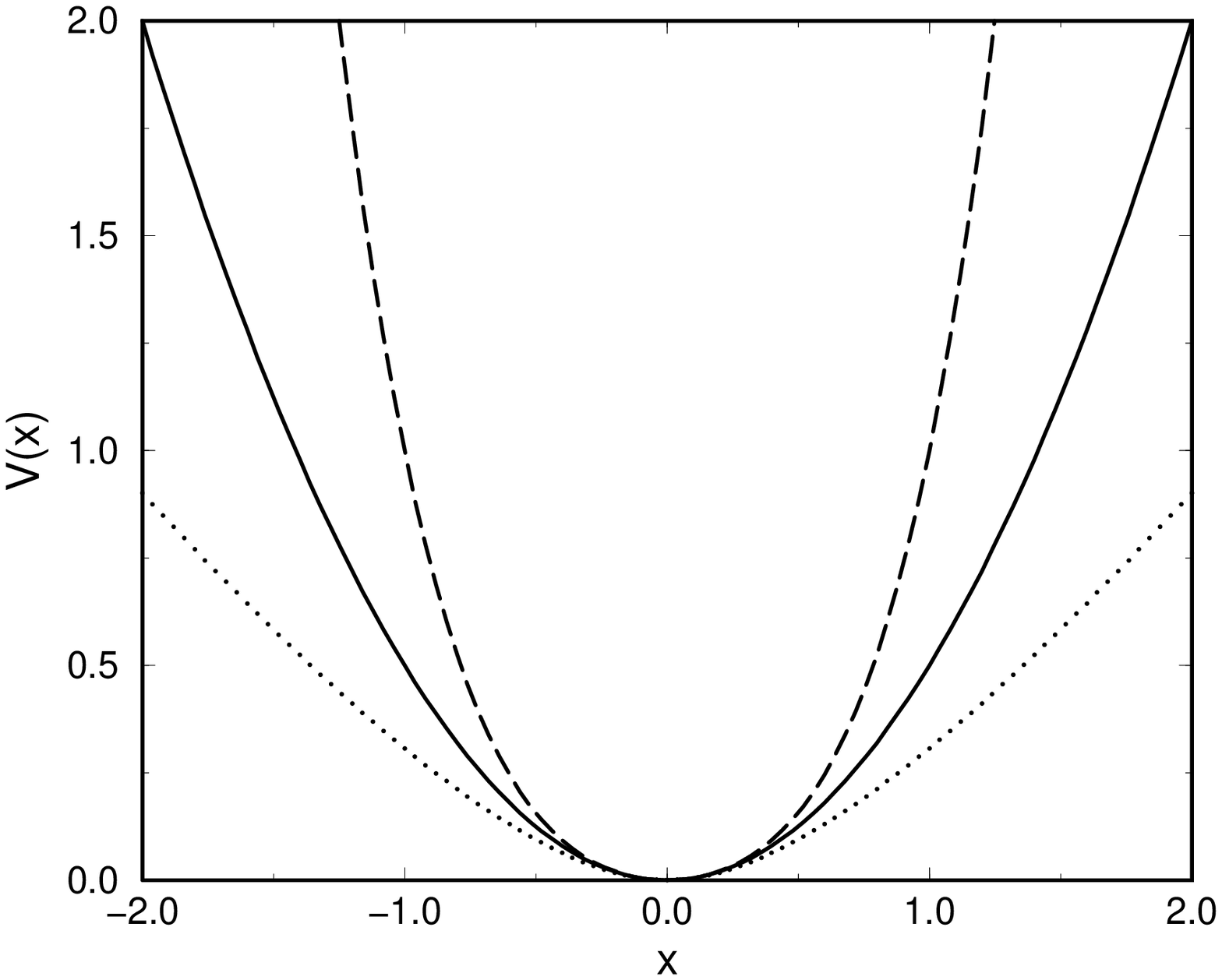}
\leavevmode
\epsfxsize = 3.0in
\epsffile{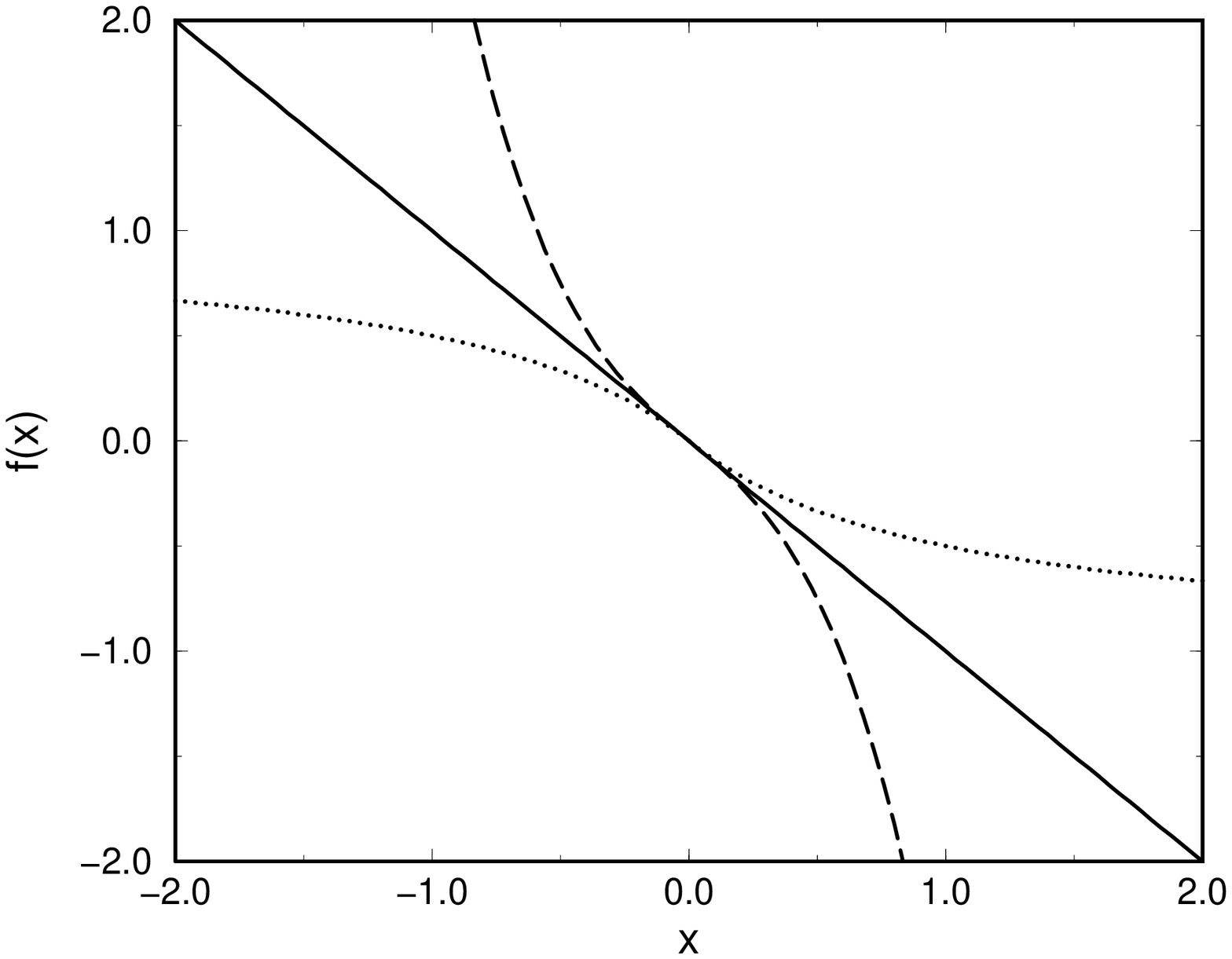}
\vspace{-0.2in}
\end{center}
\caption
{Left panel: the on-site potentials defined in Eq.(\ref{pot}).
Right panel: the associated forces. 
Solid lines: harmonic potential, $V_0(x)$.
Dotted lines: soft anharmonic potential, $V_s(x)$.
Dashed lines: hard anharmonic potential, $V_h(x)$.} 
\label{gv}
\end{figure}

Since we are interested in assessing the effects of anharmonicities
on energy localization, we start by specifying the 
on-site potentials to be used in our analysis:
\begin{equation}
\begin{split}
V_0(x)&=\frac{1}{2}x^2 \\
V_s(x)&=|x|-ln(1+|x|) \\
V_h(x)&=\frac{1}{2}x^2+\frac{1}{2}x^4.
\end{split}
\label{pot}
\end{equation}
The subscript $0$ stands for the harmonic case, $s$ for the soft and
$h$ for the hard.
At small amplitudes the three potentials are harmonic with a
unit natural frequency. 
Fig.~\ref{gv} shows the potentials and associated forces.
 
We end this section with a brief description of the numerical methods used
in our simulations throughout this paper.
The numerical integration of the stochastic equations for all our
simulations is performed using the second order Heun's method
(which is equivalent to a second order Runge Kutta integration)
\cite{Gard,Toral}.
We use a time step $\Delta t = 0.005$. The
number of oscillators in our simulations ranges between
100 and 1000 and is indicated in each figure as appropriate.  In
each simulation the system is initially allowed to relax
for enough iterations to insure thermal equilibrium, after which we
take our ``measurements."  In all of our subsequent energy
landscape representations we have used the same sequence of random
numbers to generate the thermal fluctuations.

\section{Properties of Uncoupled Oscillators: Entropic Localization}
\label{statloc}

In order to understand the equilibrium properties of a chain of oscillators
it is useful to first review the behavior of single (uncoupled) oscillators
described by the potentials in Eq.~(\ref{pot}). 

Suppose first that our oscillator
is {\em isolated}.  The salient features of anharmonic oscillators 
are that 1) they oscillate with different frequencies
at different energies, and 2) the density of states 
changes with changing energy. In particular, hard potentials are
associated with increasing frequencies of oscillation and sparser densities
of states with increasing amplitude (energy); on the other hand,
in soft potentials the oscillation frequency decreases and the
density of states increases with increasing amplitude.  

\begin{figure}[htb]
\begin{center}
\leavevmode
\epsfxsize = 3.in
\epsffile{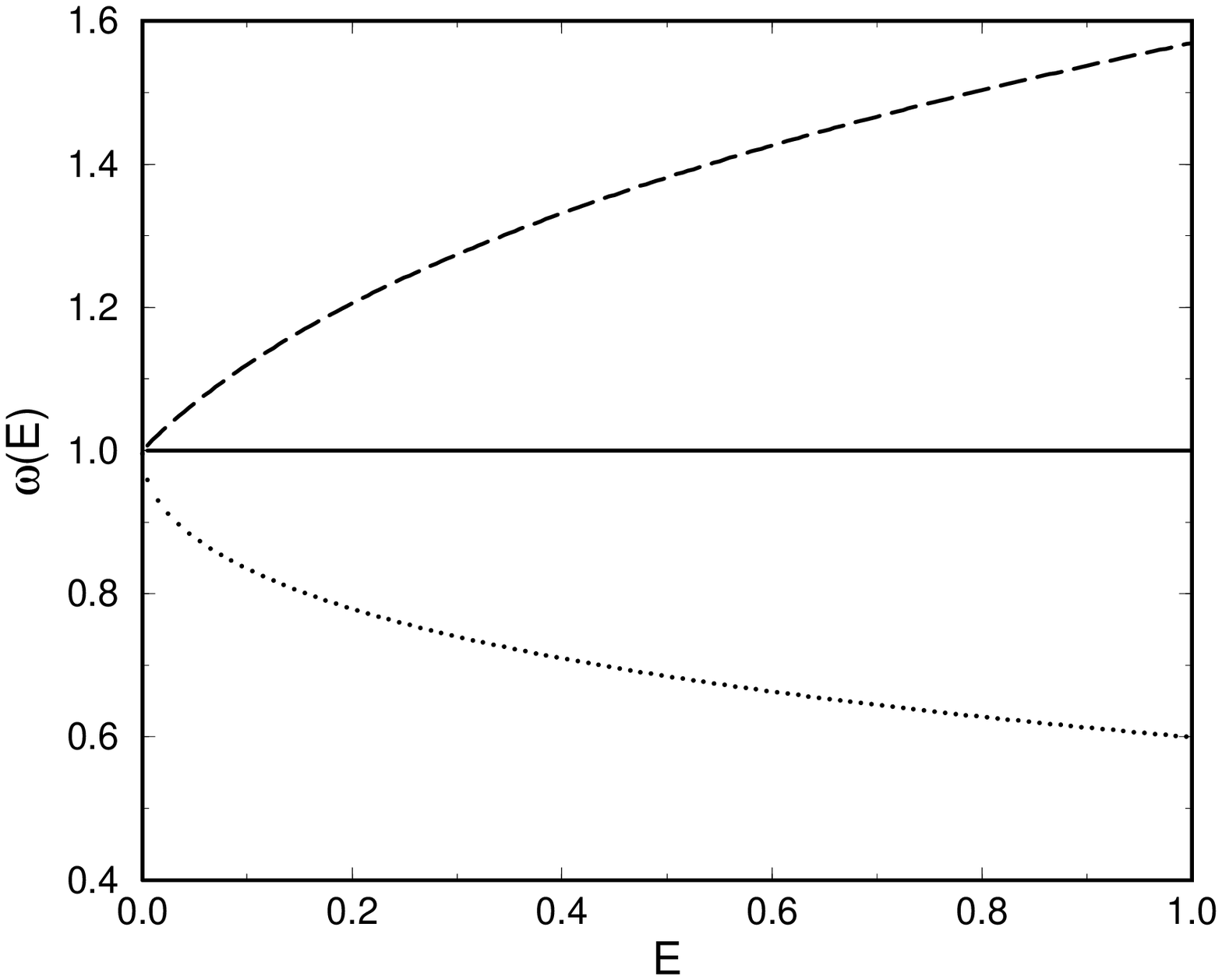}
\leavevmode
\epsfxsize = 3.in
\epsffile{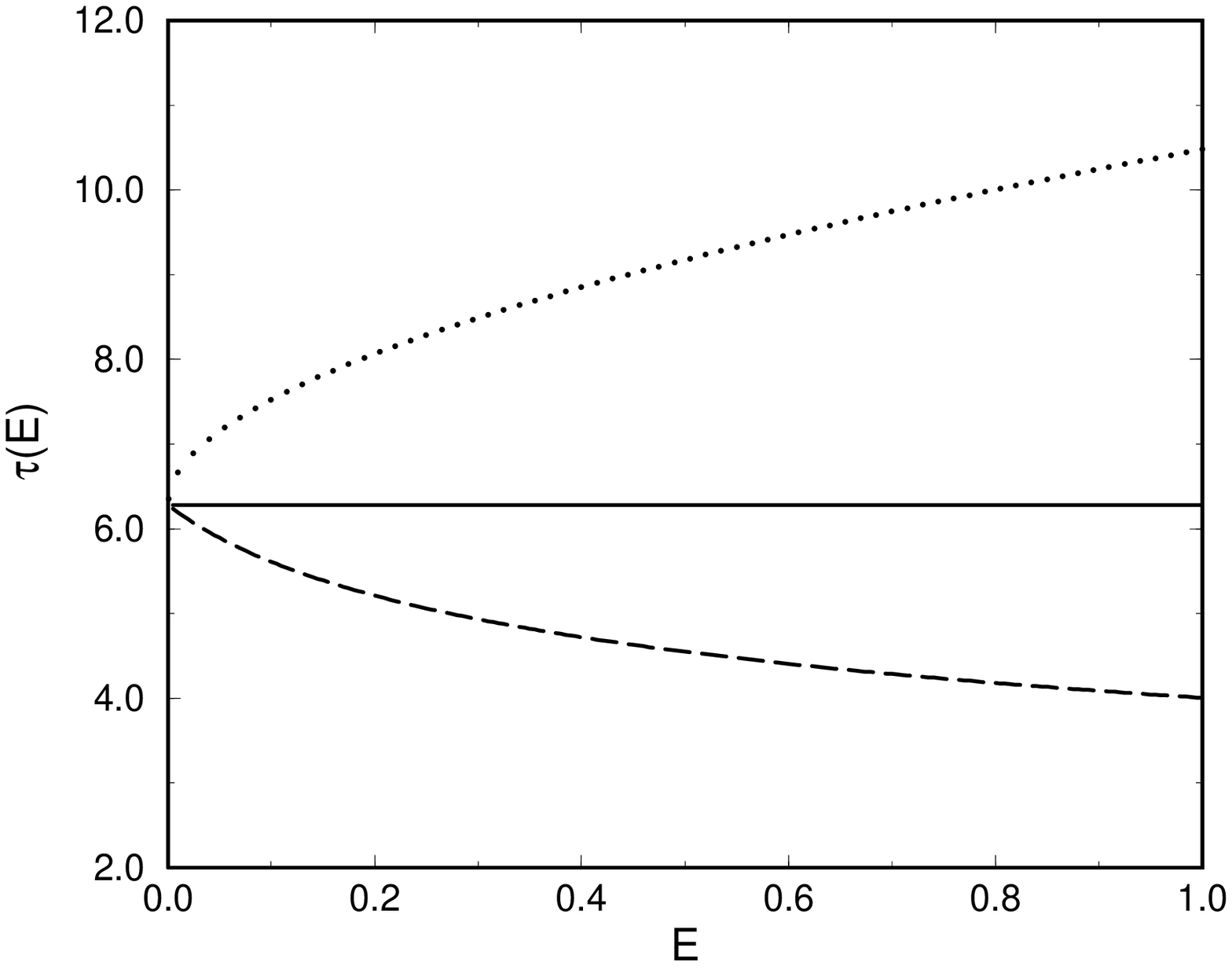}
\vspace{-0.2in}
\end{center}
\caption
{Oscillation characteristics of single isolated oscillators. Left 
panel: frequency
as a function of the oscillator energy for the potentials in
Eq.~(\ref{pot}).  Right panel: oscillation periods for single oscillators.
Solid lines: harmonic oscillator.
Dotted lines: soft anharmonic oscillator.
Dashed lines: hard anharmonic oscillator.
}
\label{gfr}
\end{figure}

To get a sense, useful for later analysis, of these and associated
oscillator characteristics, we present several figures that show
various distinct features of our three types of oscillators. 
Figure~\ref{gfr} shows the frequencies $\omega(E)$ of
{\em isolated single} oscillators as a function of increasing energy $E$
(which in turn corresponds to increasing
amplitude).  This frequency is evaluated directly by solving the
equation of motion $dx/dt = \pm\sqrt{2[E-V(x)]}$ over one period
of oscillation at energy $E$:
\begin{equation}
\omega(E) = \pi
\left(\int_{-x_{max}}^{x_{max}} \frac{dx}{\sqrt{2[E-V(x)]}}\right)^{-1}~.
\label{frequency}
\end{equation}
The amplitude of oscillation $x_{max}$ at a given energy
can be found by solving for the positive root of $V(x)=E$.
The harmonic oscillator has a
single frequency at unity.
The soft and hard oscillators oscillate at unit
frequency at low amplitudes (energies) because we have chosen all the
oscillators to coincide there, but with increasing amplitude the hard
oscillator frequencies increase and those of the soft oscillator decrease.
In Fig.~\ref{gfr} we also show the period of oscillations
$\tau(E)=2\pi/\omega(E)$.  The period increases with increasing
energy for the soft oscillator, remains constant for the harmonic
oscillator, and decreases with energy for a hard oscillator.  This behavior
will figure prominently in our subsequent analysis of energy localization. 

\begin{figure}[htb]
\begin{center}
\leavevmode
\epsfxsize = 3.0in
\epsffile{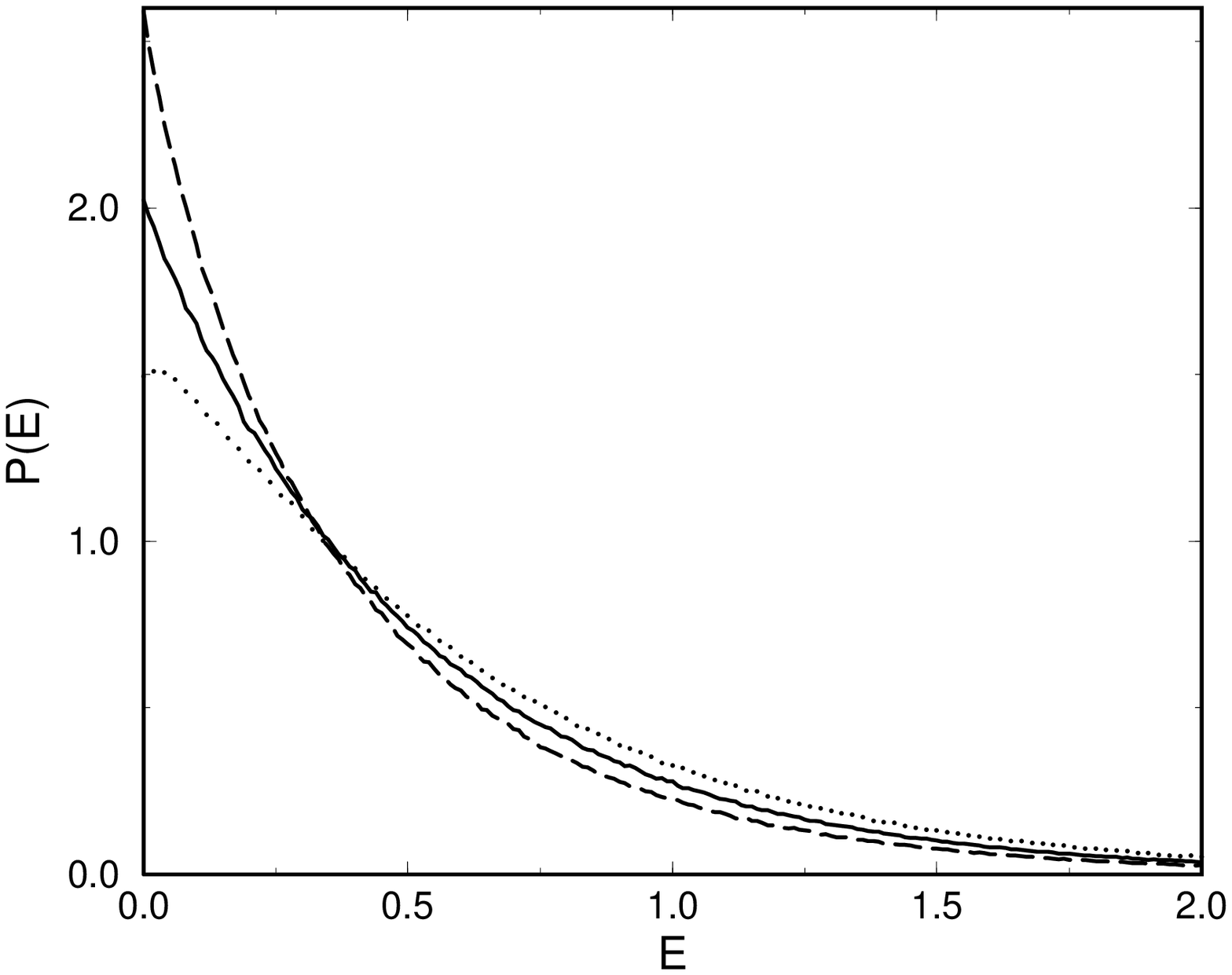}
\leavevmode
\epsfxsize = 3.0in
\epsffile{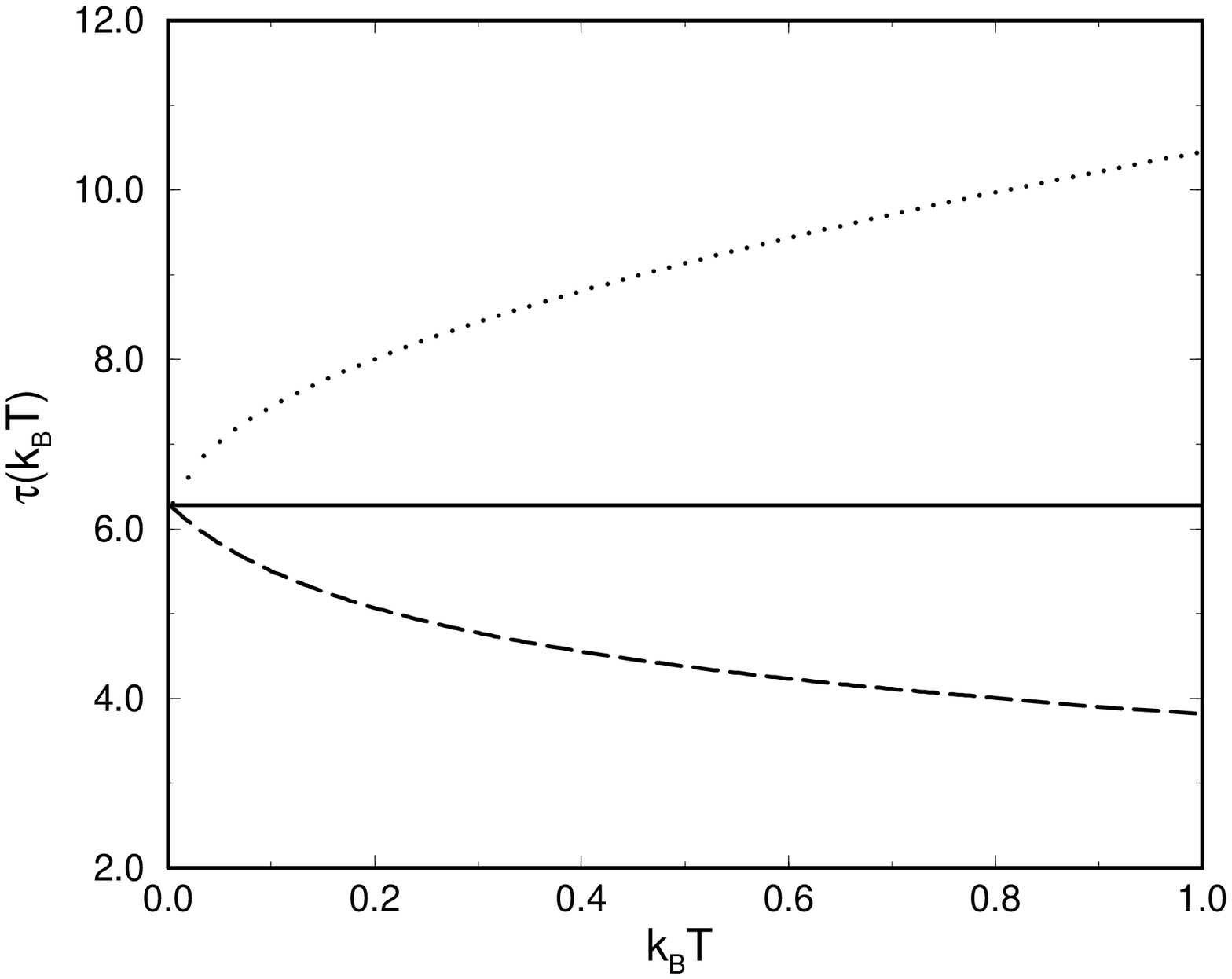}
\vspace{-0.2in}
\end{center}
\caption
{Left panel: energy distribution in single thermalized oscillators
for the three potentials at $k_BT=0.5$. Right panel: average oscillation
period for the three oscillators as a function of temperature.
Solid lines: harmonic potential, $V_0(x)$.
Dotted lines: soft anharmonic potential, $V_s(x)$.
Dashed lines: hard anharmonic potential, $V_h(x)$.  
}
\label{pvse}
\end{figure}

Next we consider these same single oscillators, but now each connected
to a heat bath at temperature $T$ via Langevin terms.  The left panel of
Fig.~\ref{pvse} shows the normalized
energy distribution $P(E)$ {\em vs} $E$ for the three cases.
This distribution is given by
\begin{equation}
P(E)=\frac{e^{-E/k_BT} \tau(E)}{\int_0^\infty dE e^{-E/k_BT} \tau(E)}
\label{distribution}
\end{equation}
where the density of states is just the period of oscillations.
The figure supports
our introductory comments: firstly, that the average energy
of the soft oscillators is greater than that of the harmonic oscillators,
whose average energy is in turn greater than that of the hard
oscillators; 
secondly, that the  energy fluctuations are smallest in the hard
oscillator and largest in the soft
oscillator.
Thus in equilibrium we find at any instant that there is a greater
variability of energy in an ensemble of single soft oscillators than
in one of harmonic or hard oscillators.   The right panel of
Fig.~\ref{pvse} shows the average period of oscillation $\tau(k_BT)$ for
a thermalized distribution:
\begin{equation}
\tau(k_BT)\equiv \int_0^\infty dE~\tau(E) P(E)~.
\label{period}
\end{equation}
Consonant with the energy dependence of $\tau(E)$, the average period of
the soft oscillator increases with temperature, that of the harmonic
oscillator is independent of temperature, and that of the hard oscillator
decreases with temperature.

\begin{figure}[ht]
\begin{center}
\epsfxsize = 4.in
\epsffile{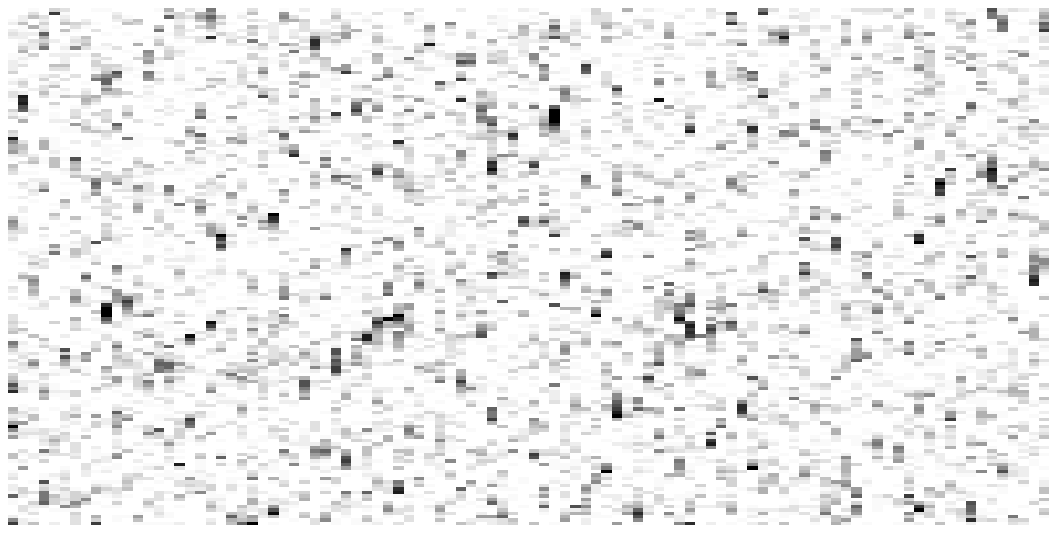}
\vspace{0.1in}
\epsfxsize = 4.in
\epsffile{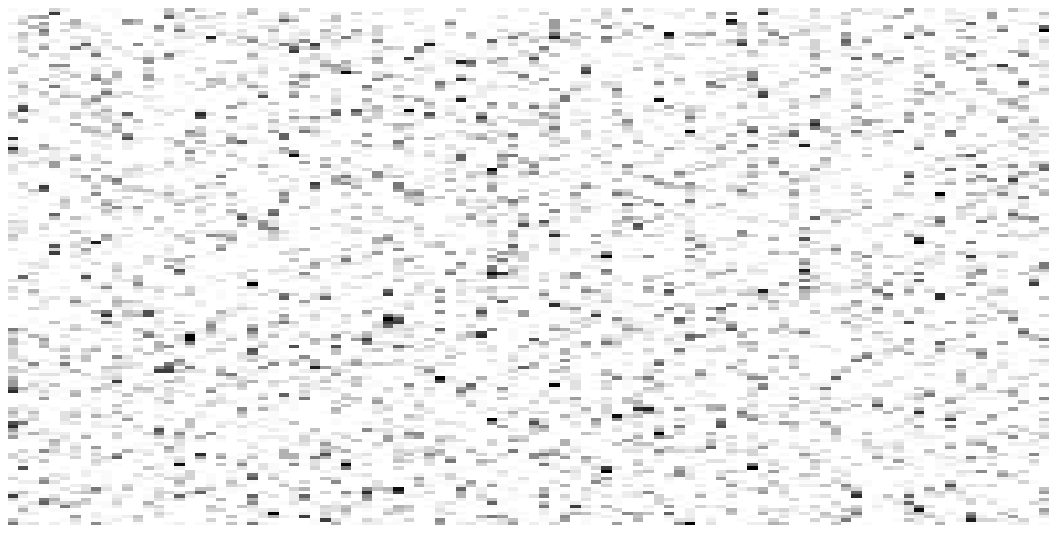}
\vspace{0.1in}
\epsfxsize = 4.in
\epsffile{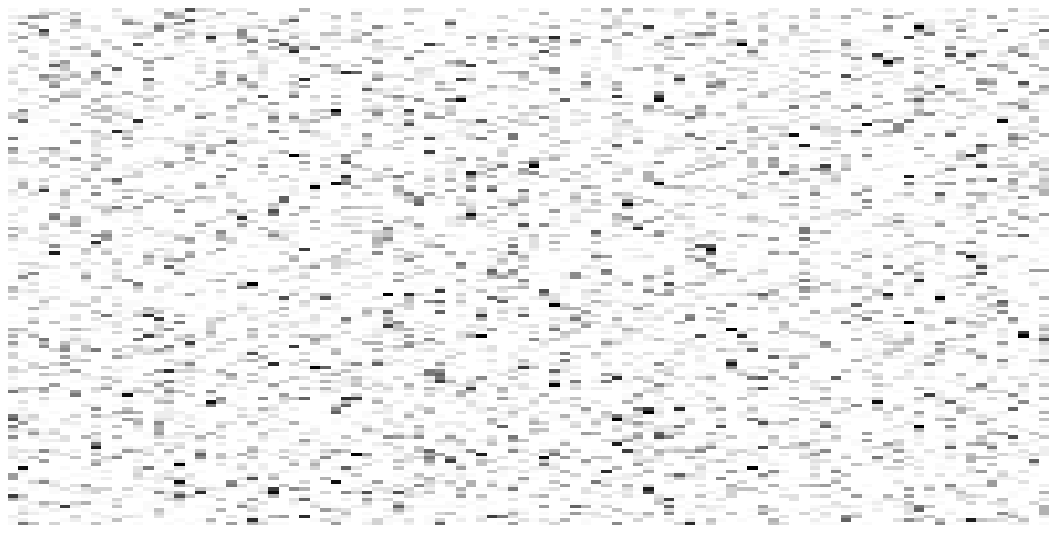}
\vspace{-0.3in}
\end{center}
\caption
{Energy (in grey scales) for ensembles of $100$ thermalized independent
oscillators as a function of time.  The oscillators are lined up (but not
connected) along the $x$-axis and time advances along the $y$-axis.  The
temperature is $k_BT=0.5$ and the dissipation parameter is $\gamma=1$.
Top panel: soft oscillators; middle panel: harmonic oscillators; lower
panel: hard oscillators.}
\label{surf2}
\end{figure}

The features just discussed are also visible in the energy landscape
rendition shown in Fig.~\ref{surf2}.  Along the horizontal direction
in each panel
lies an ensemble of 100 independent thermalized oscillators and the
vertical upward progression shows how these oscillators evolve with
time in the
equilibrium system. 
Here and in all our energy landscape figures the $y$ axis covers 120 time
units, the same units shown on time axes throughout the paper.  
Each oscillator is connected to a heat bath.
The grey scale represents the energy -- an oscillator
of higher energy is darker in this portrayal.  

The first thing to note is
that along any horizontal line (i.e. at any given time) the soft landscape
is darker and grainier than the harmonic, and the lightest and least grainy
is the hard oscillator landscape.  This reflects the fact that the
soft oscillators have the highest energies and the greatest energy
fluctuations.  
This observation provides a basis to be used in answer to the first
question posed in the introduction.  In an ensemble of independent
oscillators in thermal equilibrium there is of course a greater energy
in some oscillators than in others simply because there are energy
fluctuations in a system in thermal equilibrium.  These fluctuations are
greater in soft anharmonic oscillators than in harmonic or hard anharmonic
oscillators.
\clearpage

\clearpage

\begin{figure}[htb]
\begin{center}
\hspace{4.in}
\epsfxsize = 3.9in
\epsffile{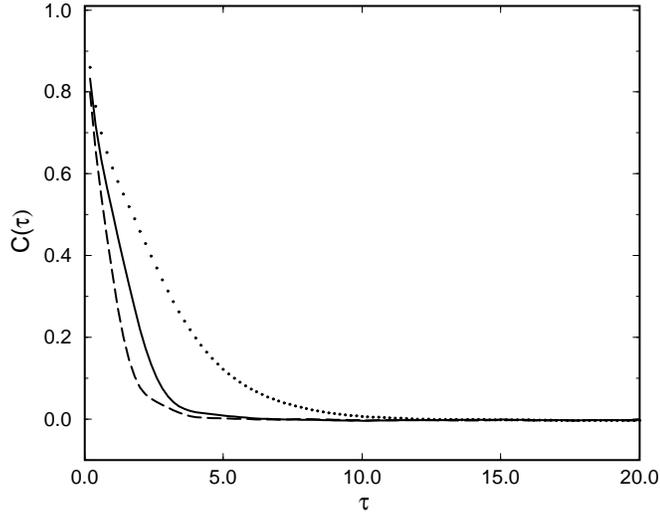}
\end{center}
\vspace{-0.3in}
\caption
{Energy correlation function {\em vs} time for independent oscillators with
$k_BT=0.5$ and $\gamma=1$.
Note that the energy changes most slowly in the soft potential ensemble.
Solid line: harmonic potential.
Dotted line: soft anharmonic potential.
Dashed line: hard anharmonic potential.}
\label{g12}
\end{figure}

The second noteworthy feature of the landscape
illustrates the answer to the second
question posed in the introduction, namely, how long it takes in an
equilibrium ensemble for the fluctuations to relax and the energy landscape
to change.  The trend for our independent oscillators is clear: the soft
oscillators retain a given energy for a longer time than do the harmonic,
which in turn hold on to a given energy longer than do the hard oscillators.
This is particularly evident for those oscillators that
acquire a high energy through a fluctuation: in the soft oscillator
landscape the dark streaks are clearly visible.  The reason for this
behavior becomes clear if we write the equation of motion for the energy
$E=p^2/2 +V(x)$ for each oscillator.  Setting $p=\dot{x}$ and using
Eq.~(\ref{lang}) one finds that for any type of oscillator
\begin{equation}
\dot{E} = -\gamma p^2  +p\eta(t).
\label{lange}
\end{equation}
Thus, the energy exchange with the surroundings involves only the
{\em momentum} variable (i.e., the kinetic energy). 
Consider an oscillator that has acquired a given high-energy fluctuation
$E$, and consider how this energy is distributed between the oscillator
displacement and momentum.
In a harmonic oscillator the energy during one cycle of
oscillation is equally partitioned between kinetic and potential.  In a
soft oscillator, however, the energy spends relatively more time in
potential than in kinetic form (and the opposite is true for the hard
oscillator). Thus, during the major portion of the cycle the momentum
of a soft oscillator is relatively low (while its displacement is large);
the energy in the soft oscillator can therefore not enter from
and leave to the thermal surroundings
as easily as in the other oscillators.  The energy relaxation process is
therefore slower, and a soft anharmonic oscillator retains a high
energy it might have gained via a fluctuation for a longer time~\cite{Brown}.

The energy relaxation process visible in Fig.~\ref{surf2} is shown more
quantitatively in Fig.~\ref{g12}.  Here we have plotted the normalized
energy correlation function
\begin{equation}
C(\tau) =  \left< \frac{\langle E(t)E(t+\tau)\rangle - \langle
E(t)\rangle \langle E(t+\tau)\rangle }{\langle E^2(t)\rangle  - \langle
E(t)\rangle^2}\right> .
\label{cor}
\end{equation}
The inner brackets indicate an average over time $t$ (200,000 iterations)
and the outer brackets an average over an ensemble of 1000 oscillators.
The correlation function is normalized so that all energies, high and low,
contribute ``equally."  It is thus a measure of the full exchange of 
energy with the heat bath, both through the dissipative term and also via
the fluctuations.
We note that the trend in Fig.~\ref{g12} (slower decay as the oscillators
soften) is consistent with the corresponding slowing trend for each
temperature in the right panel of Fig.~\ref{pvse}.  Also note that on
average the energy of an oscillator changes on the time scale of half a
period of oscillation, i.e. on the time scale it takes the oscillator to
move from one side of the potential well to the other.

We have thus summarized and illustrated our earlier
findings~\cite{Brown}, namely, that
in an array of independent oscillators in thermal equilibrium
at a given temperature there
are larger energy fluctuations and longer retention of energy
the softer the oscillators.  This is an entropy-driven localization,
arising from the fact that the density of states in soft oscillators
increases with increasing energy.  It minimizes the free energy because
it is entropically favorable
for oscillators to populate phase space regions where the density of
states is higher, which in an ensemble of soft oscillators leads to a
greater spatial variability than in harmonic or hard oscillators.  The
temporal persistence of this greater variability is a consequence of the
fact that coupling to a heat bath occurs only via the kinetic energy.
In the soft ensemble the energy is in potential form a greater fraction
of time than in kinetic form, which is not the case for the other
ensembles.  
\clearpage

\begin{figure}[htb]
\begin{center}
\epsfxsize = 4.in
\epsffile{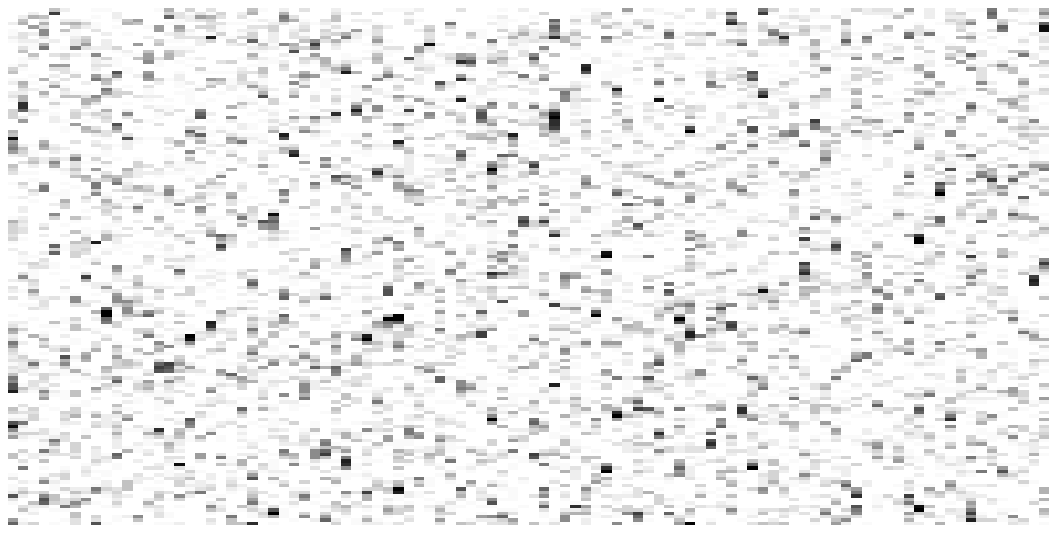}
\vspace{0.05in}
\epsfxsize = 4.in
\epsffile{surfsoft05.eps}
\vspace{0.05in}
\epsfxsize = 4.in
\epsffile{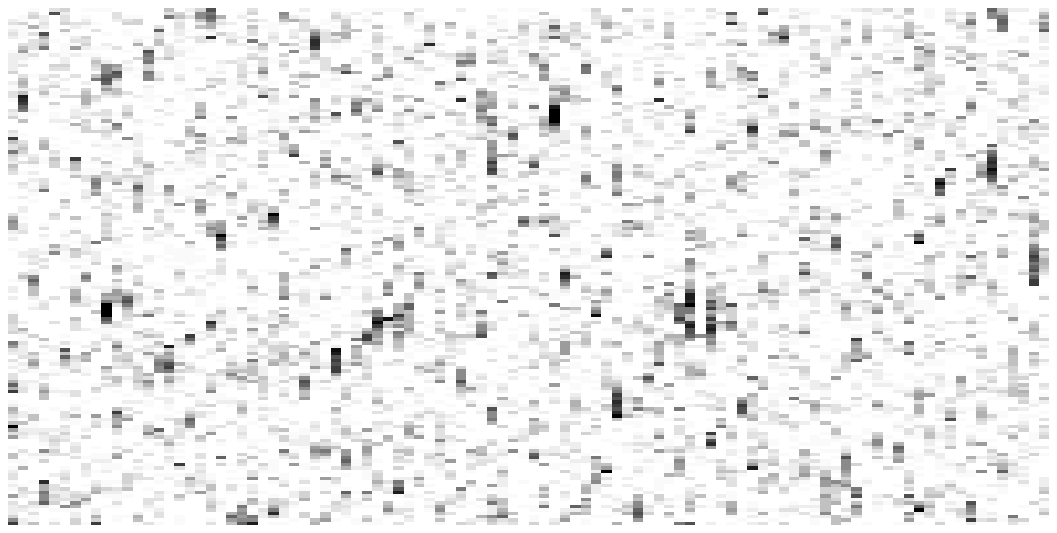}
\vspace{0.05in}
\epsfxsize = 4.in
\epsffile{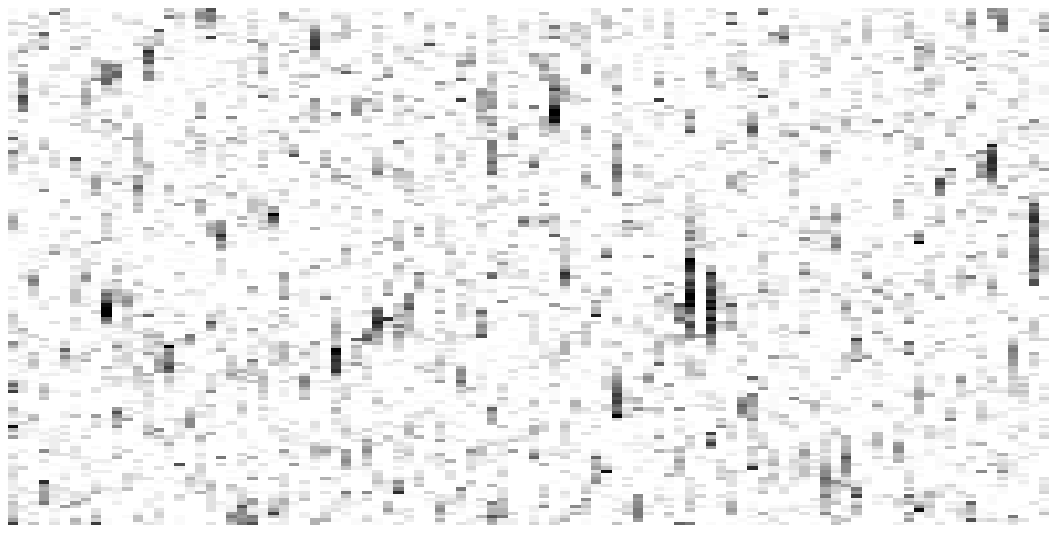}
\vspace{-0.3in}
\end{center}
\caption
{Energy landscapes for thermalized independent soft
oscillators as a function of time for different temperatures.
The dissipation parameter is $\gamma=1$. Temperatures 
from top to bottom:  $k_BT=0.1,~0.5,~1.0$ and $2.0$.}
\label{surfsofteps}
\end{figure}
\clearpage

We gave this scenario the name stochastic localization in our earlier
work~\cite{Brown}, but will refer to it as {\em entropic localization},
a term that more accurately reflects its physical causes.  It is
important to
stress that entropic localization in soft oscillators is robust in the
sense that it becomes {\em more} pronounced as temperature increases
provided the potential continues to soften, 
and that it is achieved regardless of the initial
condition of the system.  

The remaining parameters that can be varied at this point are the
{\em dissipation parameter} and the {\em
temperature}. 
A change in the dissipation
parameter does not affect Fig.~\ref{pvse} since this is an equilibrium
distribution. 
In Fig.~\ref{surf2} a higher dissipation parameter would cause
a more rapid decay of energy fluctuations (and, correspondingly,
a lower dissipation parameter allows an energy fluctuation to survive for a
longer time). Thus, although high dissipation does not interfere with
the appearance of greater energy fluctuations in the soft oscillators,
it works against the temporal retention of excess energy by any one
oscillator.  The energy correlation function decays 
more slowly for the soft oscillator for any dissipation, and 
this decay is more rapid (for all the oscillators) as the dissipation
increases. In any case, for a given dissipation parameter
the softer potential retains energy for a longer time.

\begin{figure}[ht]
\begin{center}
\hspace{4.in}
\epsfxsize = 4.0in
\epsffile{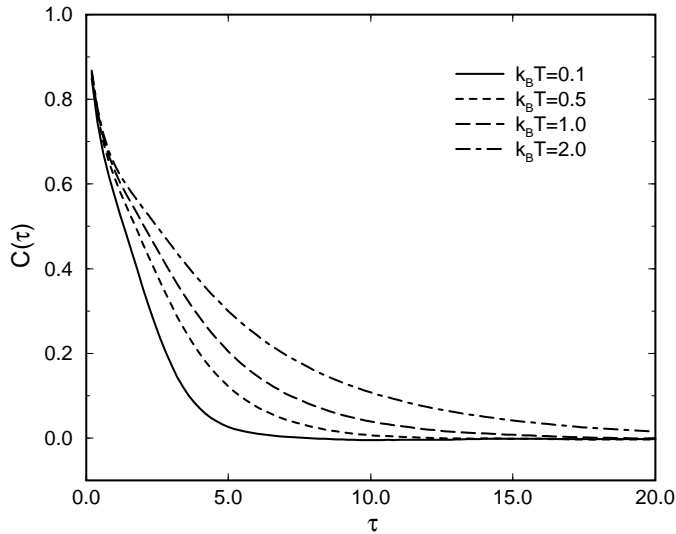}
\vspace{-0.2in}
\end{center}
\caption
{Energy correlation function {\em vs}
time for independent soft oscillators with
$\gamma=1$ and different temperatures (the same as in Fig.~\ref{surfsofteps}).}
\label{corsofteps}
\end{figure}

The temperature affects the quantitative outcome of Figs.~\ref{pvse}
and \ref{surf2}.  In Fig.~\ref{pvse} the distributions 
broaden with increasing temperature, but the differences between the
different oscillators remain and, in particular, the fact that the
distribution for the soft oscillator is the broadest continues to be true.
In Fig.~\ref{surf2} higher temperatures produce relatively greater
graininess in the soft oscillator figure than in the other two.
This is clearly observed in the sequence of
Fig.~\ref{surfsofteps}, which shows the evolution of ensembles of
soft oscillators for different temperatures. 
A temperature increase leads to stronger entropic
localization and this effect also appears in the energy correlation
functions, as shown in
Fig.~\ref{corsofteps}.  This behavior is contrasted with that of
harmonic and hard anharmonic oscillators, whose energy landscapes and
energy correlation functions show essentially no
temperature dependence in this range.  The energy fluctuations in
these latter cases dissipate very quickly.
Note that the temperature dependence of the correlation times implicit in
Fig.~\ref{corsofteps} is consistent with the temperature dependence of
an average period of oscillation of a soft oscillator as shown in the right
panel in Fig.~\ref{pvse}: with increasing temperature the correlation time
continues to be approximately half a period.  

With this background, we are now ready to consider the behavior of {\em
chains} of oscillators, where everything that we have found so far has to
be reconsidered in the face of the additional forces now present through
the oscillator--oscillator coupling.

\section{Coupled Oscillators}
\label{coupled}
In this section we explore the consequences of coupling the oscillators
discussed in the previous section with harmonic springs.  In this
exploration we attempt to bring some order to seemingly contradictory
reports that the coupled oscillators must be hard in order for such
an array to localize energy effectively, or that the coupled oscillators
must be soft in order to accomplish such localization. 
To anticipate our results: we will show that both claims are correct,
but each in a different parameter regime and for different physical
reasons.  The variable parameters
in this discussion are the {\em temperature} $k_BT$, the
{\em dissipation parameter} $\gamma$, and the
{\em coupling strength} $k$.  

In order to determine the conditions that may lead to energy localization
in a thermalized chain of oscillators it is useful to investigate
the ways in which energy may escape from a given oscillator.  It is
apparent from the Langevin equation (\ref{lang})
that there are now two channels of
escape. As in the last section, one is the friction term that
dissipates the energy to the bath.  The other is the coupling term
that transfers energy to the nearest neighbors. The difference
between these two mechanisms is that the dissipation is determined entirely
by the kinetic energy of the oscillator. Energy transfer along the chain,
on the other hand, while still dependent on the kinetic energy, is primarily
determined by the extension or contraction
of the springs connecting neighboring oscillators, that is,
by the potential energy through the relative oscillator displacements.  
To make these statements more quantitative, it is useful to generalize
the concept of a local energy by defining a local function  
whose sum over sites is the total energy of the
chain.  To include the contribution from the nearest-neighbor restoring
forces one writes
\begin{equation}
E_i\equiv \frac{p_i^2}{2} +V(x_i)+\frac{k}{4}[(x_{i}-x_{i+1})^2 +
(x_{i}-x_{i-1})^2],
\label{localenergy}
\end{equation}
and the total energy of the system is then $E=\sum_i E_i$.
The rate of change of the local energy is easily found to be 
\begin{equation}
\dot{E_i}=-\gamma p_i^2 +p_i\eta_i(t)-\frac{k}{2}(x_i-x_{i+1})(p_i+p_{i+1}) 
-\frac{k}{2}(x_i-x_{i-1})(p_i+p_{i-1}).
\label{eloss}
\end{equation}
Note that although this expression does not explicitly involve the potential,
the rate of local energy loss of course does depend on the potential
through the displacements and momenta.  

The dynamics of the local energy 
will thus depend on the interplay of the thermal (fluctuations), 
dissipative, and intrachain forces.  In order to highlight the main
comparisons and contrasts, we frequently will juxtapose the behavior
of chains for which one or the other of the energy exchange channels
is clearly the dominant one, and in each case assess the effects of
temperature changes.

\begin{figure}[htb]
\begin{center}
\epsfxsize = 4.in
\epsffile{surfsoft05.eps}
\vspace{0.05in}
\epsfxsize = 4.in
\epsffile{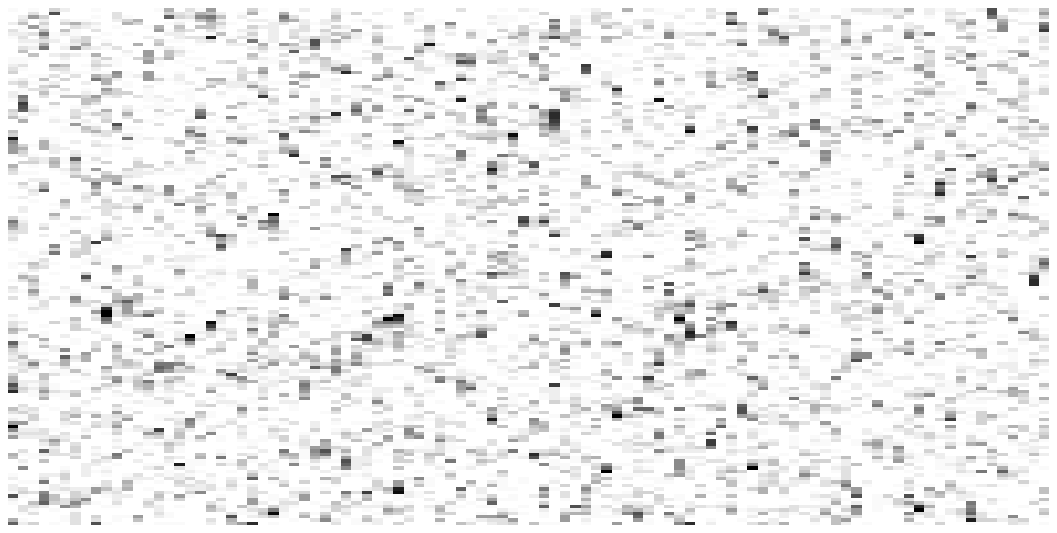}
\vspace{0.05in}
\epsfxsize = 4.in
\epsffile{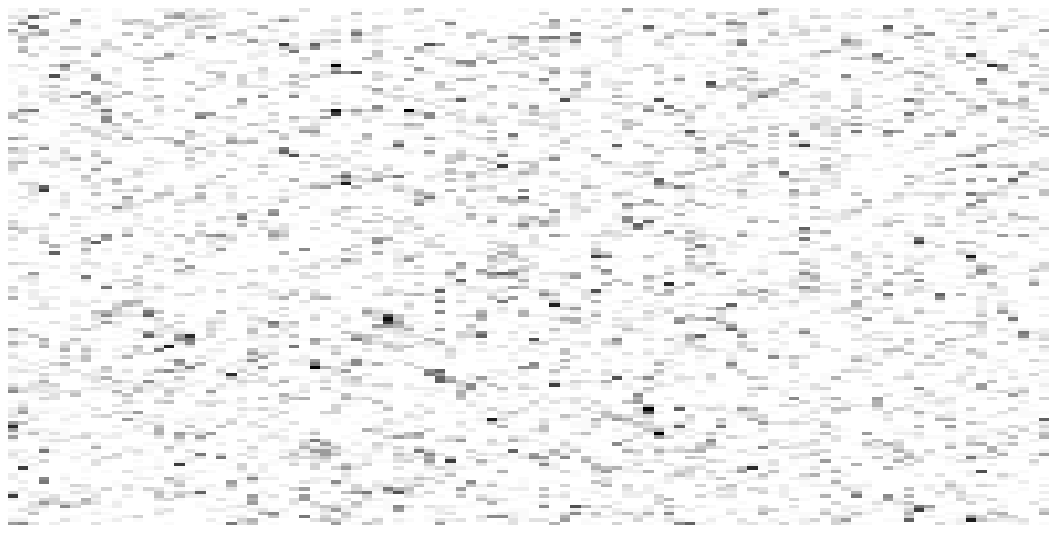}
\vspace{0.05in}
\epsfxsize = 4.in
\epsffile{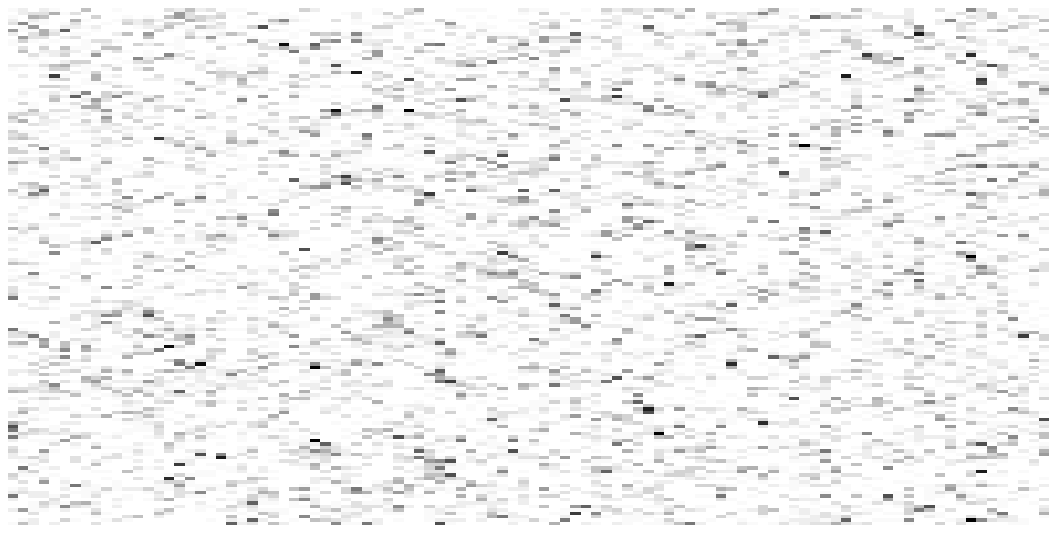}
\vspace{-0.3in}
\end{center}
\caption
{Energy landscapes for thermalized soft
oscillators as a function of time.
The dissipation parameter is $\gamma=1$ and the temperature
$k_BT=0.5$. From top to bottom the coupling constants are
$k=0, ~0.05,~ 0.5$ and $1.0$.}
\label{surfsoftk}
\end{figure}

The effect of interoscillator coupling on entropic localization
is illustrated in Fig.~\ref{surfsoftk}. In this figure we show the
system of soft oscillators
that were uncoupled in Fig.~\ref{surfsofteps} (specifically, the case
with $k_BT=0.5$ and $\gamma=1$), but now providing successively larger
values for the coupling constant $k$.
Entropic localization is still apparent
for small values of $k$, but as coupling increases there is clear
degradation of entropic localization.  This is to be expected since
energy exchange is sensitive to large oscillator amplitude differences
in soft oscillators.

\begin{figure}[ht]
\begin{center}
\hspace{3.9in}
\epsfxsize = 5.in
\epsffile{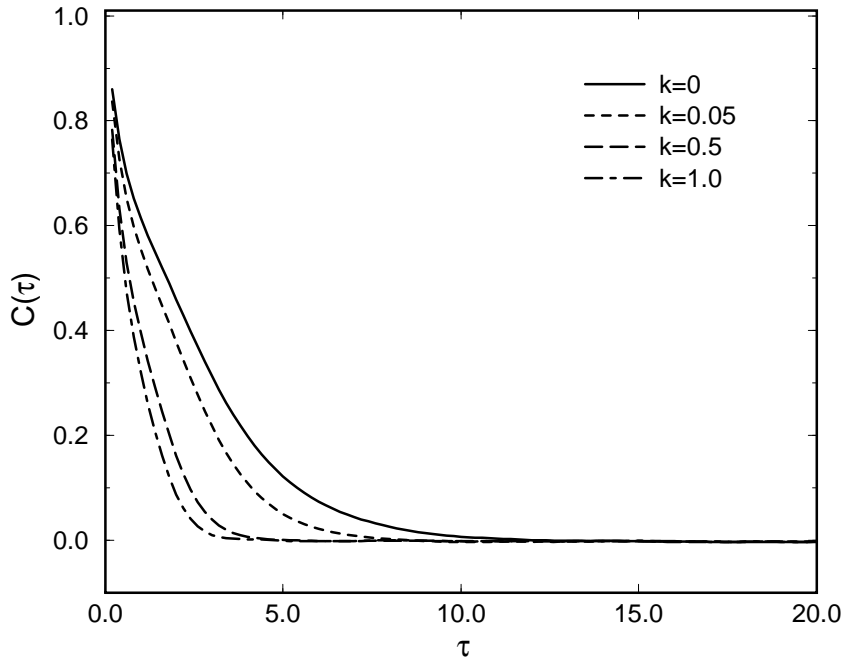}
\vspace{-0.2in}
\end{center}
\caption
{Local energy correlation function {\em vs} time for chains of soft
oscillators with
$\gamma=1$, $k_BT=0.5$ and different values for the coupling constant
(same as in Fig.~\ref{surfsoftk}).}
\label{softcork}
\end{figure}

The associated energy correlation functions
\begin{equation}
C(\tau) =  \left< \frac{\langle E_i(t)E_i(t+\tau)\rangle - \langle
E_i(t)\rangle \langle E_i(t+\tau)\rangle }{\langle E_i^2(t)\rangle  - \langle
E_i(t)\rangle^2}\right>_i
\label{cori}
\end{equation}
\clearpage
\clearpage

\noindent
for the cases in Fig.~\ref{surfsoftk} are shown in Fig.~\ref{softcork}.
These curves confirm the degradation of entropic localization
with increasing $k$.

\begin{figure}[htb]
\begin{center}
\epsfxsize = 3.9in
\epsffile{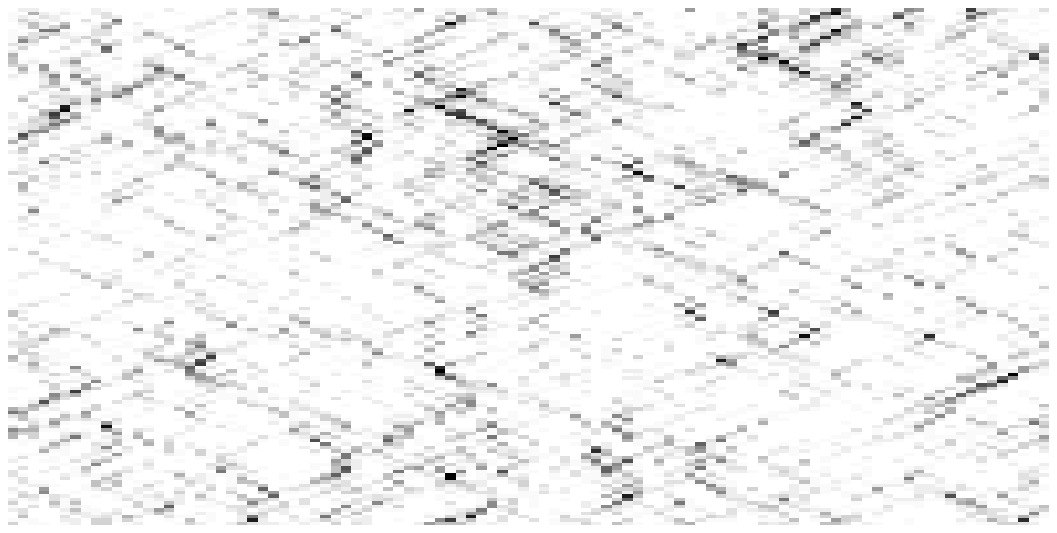}
\vspace{0.1in}
\epsfxsize = 4.in
\epsffile{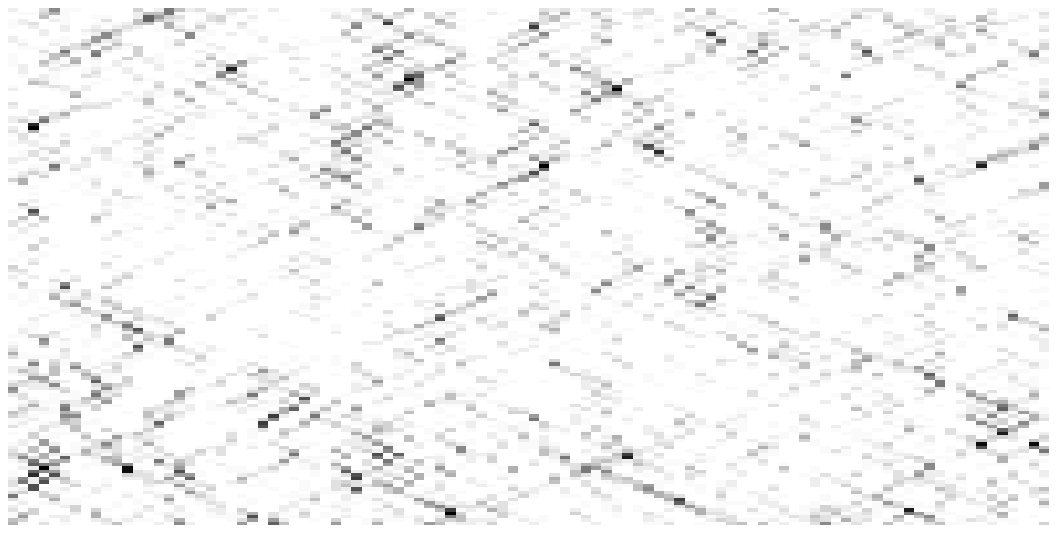}
\vspace{0.1in}
\epsfxsize = 4.in
\epsffile{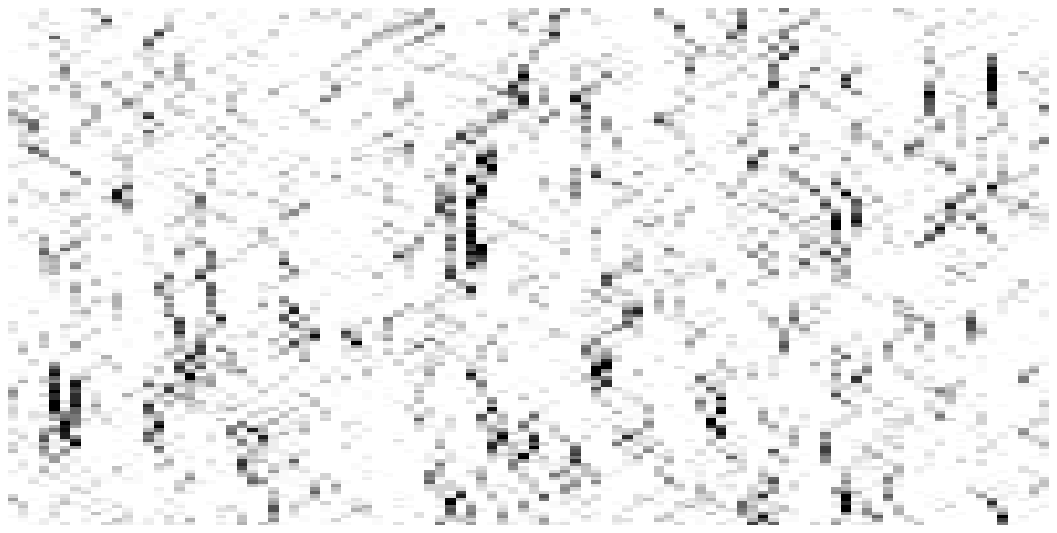}
\vspace{-0.3in}
\end{center}
\caption
{Energy landscapes for thermalized strongly coupled
oscillators as a function of time.
The dissipation parameter is $\gamma=0.05$, the temperature $k_BT=0.5$, and
the coupling constant $k=1.0$.
Top panel: soft oscillators; middle panel: harmonic oscillators; lower
panel: hard oscillators.}
\label{surfk}
\end{figure}

\begin{figure}[htb]
\begin{center}
\hspace{4.in}
\epsfxsize = 5.in
\epsffile{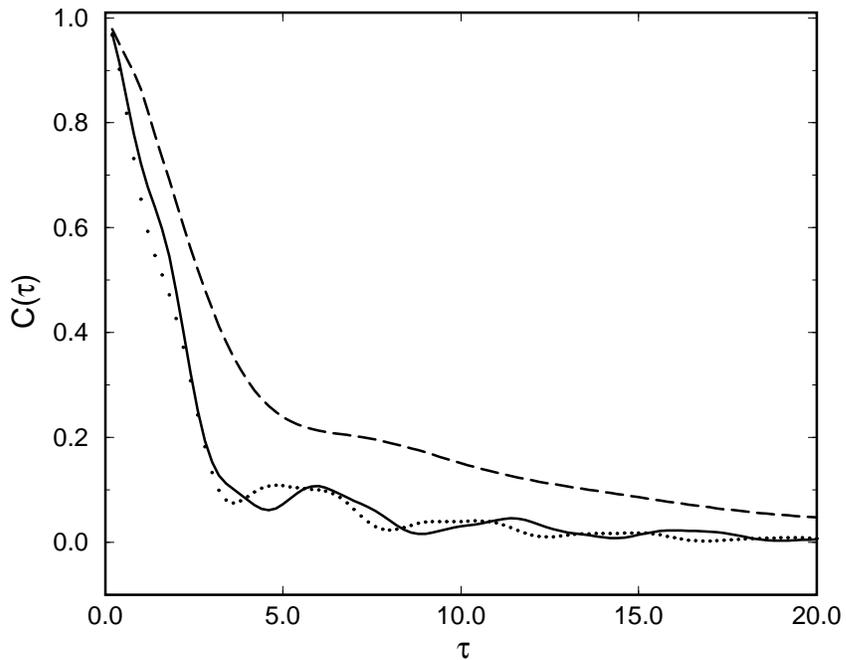}
\end{center}
\caption
{Energy correlation function {\em vs} time for coupled oscillators with
$\gamma=0.05$, $k_BT=0.5$, and $k=1.0$.
Solid line: harmonic potential.
Dotted line: soft anharmonic potential.
Dashed line: hard anharmonic potential.}
\label{cork}
\end{figure}

We thus turn to chains of coupled oscillators with low dissipation
($\gamma=0.05$) and focus, in particular, on strongly coupled chains (if
both $k$ and $\gamma$ are small we know pretty much what happens from the
analysis in the previous section).
In Fig.~\ref{surfk} we have drawn the energy landscape
for the soft (top panel), harmonic (middle panel) and hard oscillators
(lower panel)
providing $k_BT=0.5$, $\gamma=0.05$, and $k=1.0$. From this figure it
is clearly evident that now the localization of energy at a given site
is greater in the hard case
than in the harmonic case, and this in turn, is greater than in
the soft case.
The confirming local energy correlation functions for these cases are
plotted in Fig.~\ref{cork}.  Clearly, for a given temperature the hard
array retains energy at a given location for a longer time than do the
other two arrays.

In the low-$\gamma$, large-$k$ regime the effective
energy exchange channel is sensitive to the oscillator amplitude rather
than to its kinetic energy, so we expect 
entropic localization in
\clearpage
\clearpage
\noindent
the soft array 
to be degraded since soft oscillators have large amplitudes.
Furthermore, as the harmonic coupling increases it eventually overwhelms
the local soft potential and the soft chain becomes an essentially harmonic
chain at sufficiently large $k$.  
On the other
hand, hard oscillators exchange little energy via the coupling
channel since they do not reach large amplitudes. This, and the fact that
dissipation to the bath via kinetic energy (the other energy exchange
channel) has been minimized (low $\gamma$), leads to persistent energy
localization in the hard array. This is an energetic localization
mechanism.  The frequency mismatch between an energetic hard oscillator and
its less energetic neighbors, and the dearth of density of states at high
energies, further contribute to this persistence.

\begin{figure}[htb]
\begin{center}
\epsfxsize = 4.in
\epsffile{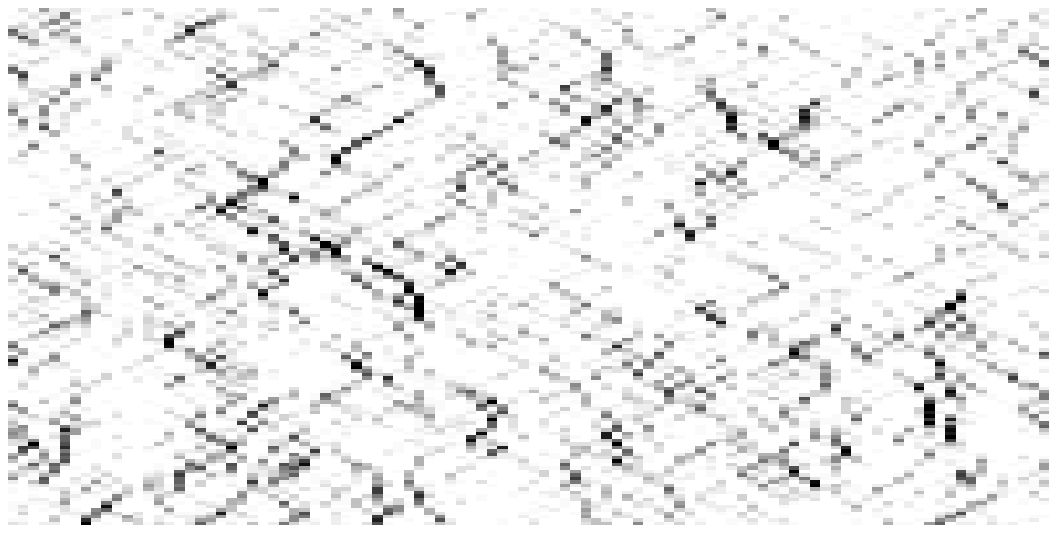}
\vspace{0.05in}
\epsfxsize = 4.in
\epsffile{surfhard05k1g005b.eps}
\vspace{0.05in}
\epsfxsize = 4.in
\epsffile{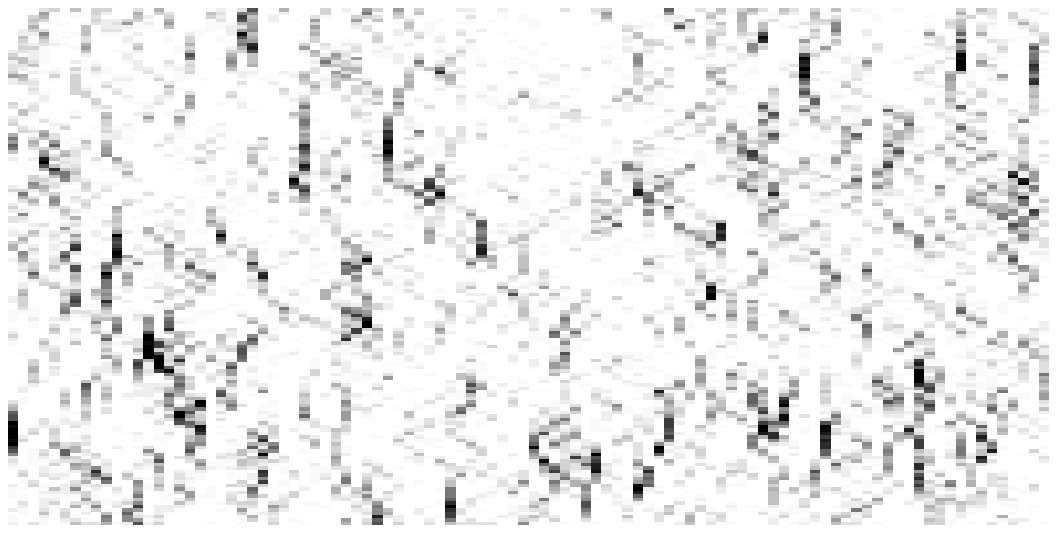}
\vspace{0.05in}
\epsfxsize = 4.in
\epsffile{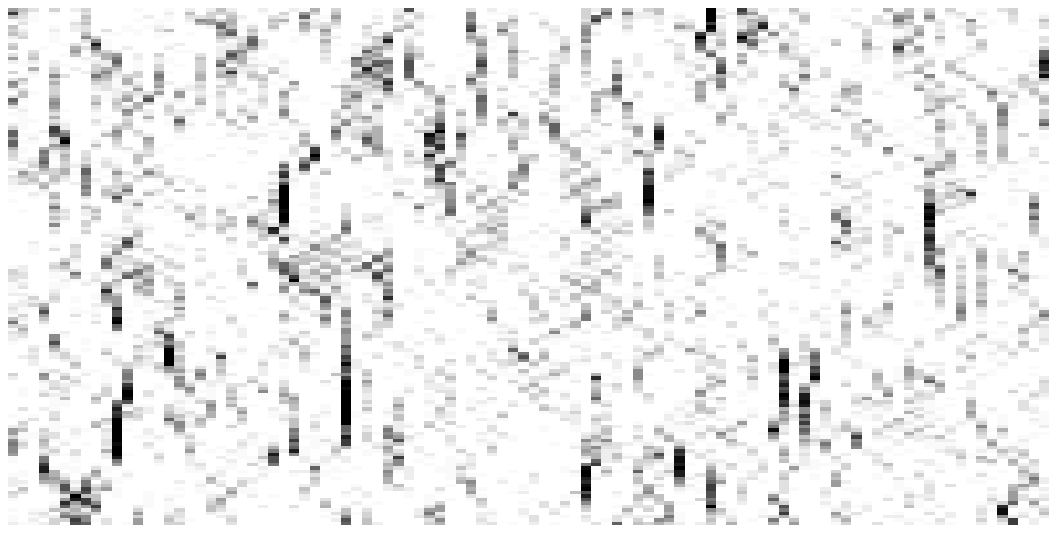}
\vspace{-0.3in}
\end{center}
\caption
{Energy landscapes for thermalized coupled hard
oscillators as a function of time.
We take $\gamma=0.05$ and $k=1.0$. Temperatures 
from top to bottom: $k_BT=0.1,~0.5,~1.0$ and $2.0$.}
\label{surfhardeps}
\end{figure}

\begin{figure}[ht]
\begin{center}
\hspace{3.9in}
\epsfxsize = 5.in
\epsffile{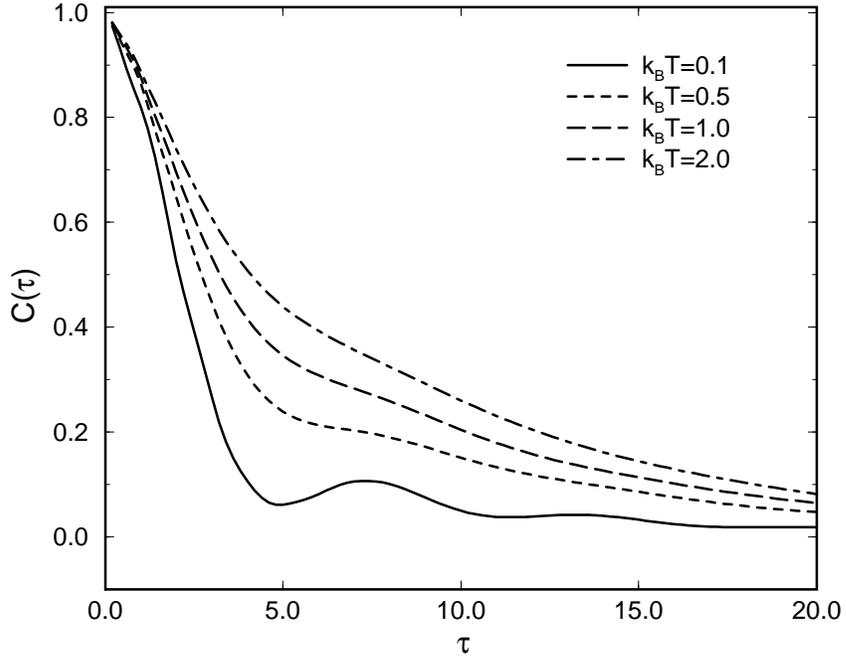}
\vspace{-0.2in}
\end{center}
\caption
{Energy correlation function {\em vs} time for strongly coupled hard
oscillators with $\gamma=0.05$, $k=1$ and different temperatures
(same as in Fig.~\ref{surfhardeps}).}
\label{corhardeps}
\end{figure}

The energetic localization mechanism in strongly coupled hard oscillators
is robust against temperature increases.  Indeed, according to our
explanation, the localization should become more
pronounced and persistent as temperature increases provided the
dissipation is sufficiently weak. 
In Fig.~\ref{surfhardeps} we have drawn the energy landscapes
for a strongly coupled
\clearpage
\noindent
($k=1.0$) array of hard
oscillators, weakly coupled to the bath ($\gamma=0.05$) at
different temperatures.
The figure qualitatively confirms these expectations.
The corresponding energy correlation functions are plotted in
Fig.~\ref{corhardeps}: $C(\tau)$ for the hard
chain does decay more slowly with increasing temperature. Thus,
localization in this strongly coupled system of hard oscillators
becomes more effective with increasing temperature and is not
entirely fragile against dissipative forces.
On the other hand, the soft and harmonic correlation
functions (not shown here) are essentially independent of temperature.
Note that the trend in Figs.~\ref{cork} and \ref{corhardeps} is
``opposite" to that of the uncoupled oscillators in the right hand panel of
Fig.~\ref{pvse}.  In the strongly coupled chain harmonic and soft
oscillators in fact lose their energy rather quickly on the time scale of
one oscillation period of an isolated oscillator, but the hard oscillators
retain energy correlations for longer than a period, indeed for many
periods at the highest temperatures shown.  With increasing temperature the
hard oscillators retain energy more effectively even while the average
oscillation period decreases.  In fact, the decay of the correlation
functions appears to involve two time scales, one of the order of an
oscillation period and another much longer one that grows with temperature.

The temporal irregularities (oscillations) visible in
Figs.~\ref{cork} and \ref{corhardeps}
are reproducibly there at all temperatures; we do not know their source.

\section{Mobility of Localized Energy Fluctuations}
\label{Transport}

The first two energy landscapes in Fig.~\ref{surfk} show what might appear
as fairly dispersionless energy transport.  Narrow high-energy pulses move
visibly along the chain before disappearing, while others appear (via
thermal fluctuations) to repeat the process elsewhere along the chain.
However, this can not be claimed to represent nonlinear behavior since the
middle panel in Fig.~\ref{surfk} in fact represents completely harmonic
system!  This serves as a cautionary note about the overinterpretation of
such results. 

\begin{figure}[!htp]
\begin{center}
\epsfxsize = 6.in
\epsffile{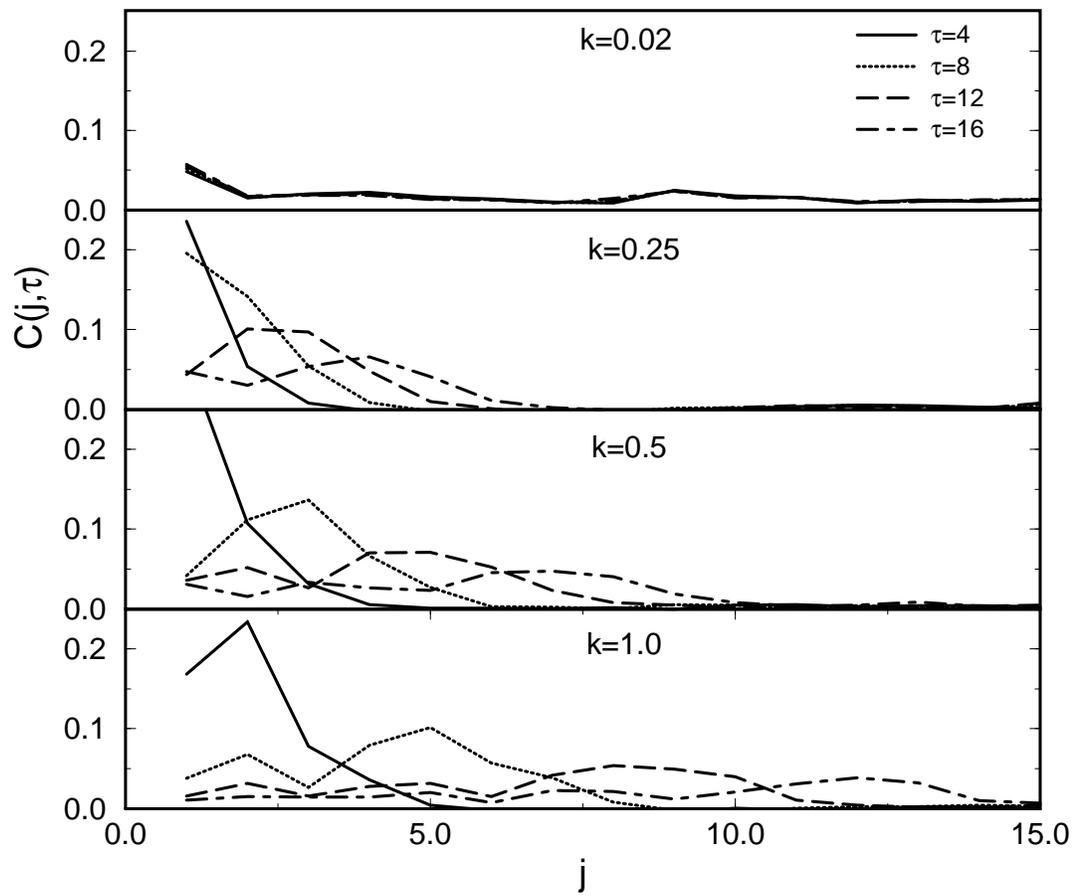}
\vspace{-0.3in}
\end{center}
\caption
{Dynamical energy correlation function $C(j,\tau)$ for
soft chains with $\gamma=0.05$ and $k_BT=1.0$.}
\label{coret1softb}
\end{figure}

\begin{figure}[!htp]
\begin{center}
\epsfxsize = 6.in
\epsffile{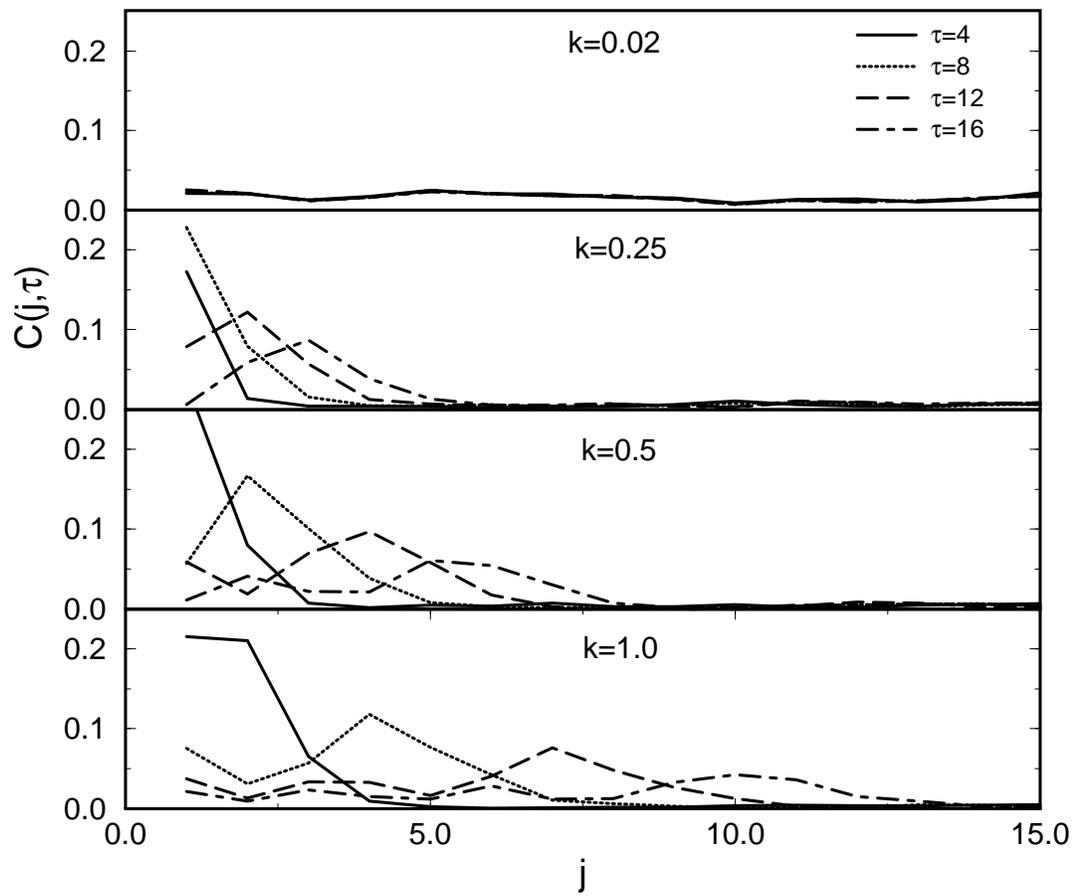}
\vspace{-0.3in}
\end{center}
\caption
{Dynamical energy correlation function $C(j,\tau)$ for
harmonic chains with $\gamma=0.05$ and $k_BT=1.0$.}
\label{coretharb}
\end{figure}

We noted earlier that with increasing $k$ the soft chain eventually becomes
essentially harmonic because the intermolecular harmonic interactions
overwhelm the local soft potential (the only way to prevent this is by
considering soft interoscillator interactions, which we defer to another
paper \cite{Sarm}).  The upper panel in Fig.~\ref{surfk} exhibits mostly this
essentially harmonic behavior -- it is quite similar to the middle panel --
but not entirely so.  The soft oscillator chain clearly shows higher-energy
regions than the harmonic (darker patches, a not fully degraded remnant of
entropic localization) that move more rapidly (steeper streaks) over longer
distances (longer streaks) than in the harmonic chain.  Therefore, the soft
anharmonicity is clearly still playing some role, albeit a diminishing one
with increasing coupling.  To provide some quantification, we introduce 
the dynamical energy correlation function
\begin{equation}
C(j,\tau) =  \left< \frac{\langle E_i(t)E_{i+j}(t+\tau)\rangle - \langle
E_i(t)\rangle \langle E_{i+j}(t+\tau)\rangle }{\langle E_i^2(t)\rangle  - \langle
E_i(t)\rangle^2}\right>_i.
\label{corjt}
\end{equation}
This correlation function plotted as a function of $j$ for various time
differences $\tau$ is shown in Fig.~\ref{coret1softb} for a soft chain
and in Fig.~\ref{coretharb} for a harmonic chain.  For a given coupling
constant $k$ and delay time $\tau$, the correlation function peaks at the
site $i+j$ to which most of the energy originally at $i$ has migrated.
The change of the peak position with $k$ indicates the velocity of the
migration, and the height and width of the pulse reflect the dispersive
dynamics.  

The following results are evident:
\begin{itemize}
\item
Increasing $k$ in either soft or harmonic chains increases the velocity at
which a fluctuation propagates.
\item
The velocity for a given set of parameters is greater in the soft chain.
\item
Dispersion is slower in the soft chain.
\end{itemize}
However, as noted before, the differences between soft and harmonic chains
at large $k$ are fairly marginal.  More dramatic differences in mobility
features occur with anharmonic intermolecular potentials, a situation that
will be presented elsewhere \cite{Sarm}.

\section{Conclusions}
\label{conclusions}
We have presented a fairly complete characterization of the thermal
equilibrium behavior of oscillator chains with ``diagonal anharmonicity,"
that is, chains with nonlinear on-site potentials and harmonic
intersite potentials.  Our particular interest lies in the characterization
of possible spatial energy localization in such systems, and of the
temporal persistence of such localization.

The instantaneous localization of energy of a system in thermal equilibrium
is a manifestation of the thermal fluctuations: it is an equilibrium
property unrelated to system dynamics.  We argued that 
not only do soft anharmonic chains have a higher total energy at a given
temperature than do harmonic or hard chains, but also that
thermal fluctuations are more pronounced in the soft anharmonic chains.
This is a consequence of the fact that free energy maximization favors
the occupation of phase space regions with a high density of states.
The density of states increases with energy in a softening potential,
so it is entropically favorable for a few soft oscillators to have rather
high energies.  This in turn leads to greater spatial energy variability
than in harmonic or hard chains, that is, soft chains have ``hotter spots." 
The effect becomes more pronounced with increasing temperature. 
This entropic energy localization mechanism in soft chains is degraded
as the harmonic intersite potential increases because the harmonic
contributions become dominant over the local soft anharmonicity effects.

In addition to the capacity for instantaneous localization of energy (which
is greatest in soft chains), one is interested in the temporal degradation
of a high energy fluctuation. That is, 
given a ``local hot spot" (which is easier to find in soft chains,
but nevertheless does occur in harmonic and hard chains due to thermal
fluctuations), how does such a fluctuation evolve in time?
Such a fluctuation never grows spontaneously, nor does it persist
indefinitely.  Rather, it eventually degrades, either through dissipation
into the bath or through dispersion along the chain.

The rate of dissipation into the bath depends on the value
of the dissipation parameter and also on the {\em kinetic} energy of
the oscillators.
If the dissipation parameter is small, this channel is of course slow for
any chain.  However, even if the dissipation parameter is large,
dissipation can still be slow if the energy is not primarily in kinetic
form.  This is the case for soft chains provided the interatomic potential
is weak (since otherwise the chain is essentially harmonic).  In soft
chains the energy is in potential form for a longer fraction of the
time than in the other chains.  As temperature is increased, this
effect becomes more pronounced because ever softer portions of
the potential become accessible, and the energy is stored as potential
energy a greater fraction of the time.

{\em Thus an increase in temperature in weakly coupled soft chains leads
not only to greater energy fluctuations but also to a
slower decay of these fluctuations.}

Energy dispersion along the chain depends on the magnitude of the
coupling constant and also on the relative oscillator displacements.
If the coupling constant is small, this channel is slow for any chain. 
If it is large, then this channel can still be slow if relative
displacements of neighboring oscillators are small.  This is the case
for the hard chain, where displacements are relatively small and
don't change much with increasing energy. 
Furthermore, because in a hard oscillator the frequency increases
with increasing energy, there is a frequency mismatch between a ``hot"
oscillator and its ``colder" neighbors that further impedes energy
transfer. This leads to greater persistence of local high-energy
fluctuations with increasing temperature.  

{\em Thus an increase in temperature in weakly dissipative hard chains leads
not only to greater energy fluctuations but also to a slower decay of
these fluctuations.}

The soft chain, on the other hand, increasingly loses its soft character
as the interoscillator energy transfer channel strengthens, and therefore
both the landscape and the dynamical effects of anharmonicity quickly
disappear as this coupling constant is increased.

Finally, we showed that in harmonically coupled nonlinear chains
(that is, in chains with ``diagonal anharmonicity") in thermal
equilibrium, high-energy fluctuation mobility does not occur beyond
that which is observed in a harmonic chain.   The situation 
might be quite different if there is ``nondiagonal anharmonicity", that
is, if the interoscillator interactions are anharmonic.  Our results
on these systems will be presented elsewhere \cite{Sarm}. 

Further presentations will also deal with bistable ``impurities"
connected to chains of the types that we have considered here \cite{Ramon}.

\section*{Acknowledgments}
The authors acknowledge helpful discussions with Dr. J. M. Sancho.
R. R. gratefully acknowledges the support of this
research by the Ministerio de Educaci\'{o}n y Cultura through
Postdoctoral Grant No. PF-98-46573147.  A. S. acknowledges sabbatical
support from DGAPA-UNAM.
This work was supported in part by the U. S. Department of Energy under
Grant No. DE-FG03-86ER13606.

\end{spacing}
\end{document}